\documentclass[10pt]{iopart}
\usepackage{multirow}
\usepackage{graphicx}
\usepackage{xcolor}
\usepackage[english]{babel}
\usepackage{diagbox}
\usepackage{hyperref}
\addto\extrasenglish{%
}
\usepackage{amsfonts}

\usepackage{subcaption}
\def\special_radius{\mathcal{R}}
\begin{document}
\title{Geodesic structure of the cosmological Levi-Civita spacetimes}
\author{Adam Tyc and Martin \v{Z}ofka}
\address{Institute of Theoretical Physics, Faculty of Mathematics and Physics, Charles University, Czech Republic}
\ead{martin.zofka@matfyz.cuni.cz}
\begin{abstract}
We investigate the implications of the behavior of geodesics in static, cylin\-drically symmetric spacetimes with a non-zero cosmological constant. We consider the symmetries of these spacetimes to restrict admissible ranges of the metric parameters and to formulate an intuitively plausible interpretation of the coordinates.
\end{abstract}

\submitto{\CQG}
\section{Introduction}
Cylindrical systems in general relativity have been studied extensively since 1919, when Tullio Levi-Civita derived the static cylindrically symmetric vacuum solution to the Einstein field equations \cite{Levi-Civita}. These systems serve both as standalone exact solutions of Einstein equations---for instance, the cosmic string \cite{Vilenkin+Shellard_2000}---and as toy models of the spacetime in the vicinity of finite prolate bodies characterized by higher multipole moments \cite{Backdahl+Herberthson_2005, Herberthson_2004}. For a recent review, see \cite{Santos+Wang_2024}. In this paper, we focus on a particular class of these solutions which are a generalization of the original solution by Levi-Civita (LC) to include the cosmological constant $\Lambda$---both positive as corresponds to inflationary cosmologies and present observations, and negative as preferred by the string theory and the AdS/CFT correspondence. The solution was found independently by Linet \cite{10.1063/1.527050} and Tian \cite{PhysRevD.33.3549} and later generalized to higher dimensions \cite{PhysRevD.79.087502,Griffiths_2010}. It is of interest that, far away from the axis of symmetry, its asymptotical behavior for $\Lambda<0$ is more closely related to that of finite bodies than is the case for zero cosmological constant---see our previous paper \cite{Zofka+Bicak_2008} where we investigated some characteristics of these spacetimes.

The generally accepted interpretation of both the LC spacetime and its gen\-er\-al\-i\-za\-tion, LC$\Lambda$, is that it describes the gravitational field outside of an infinite cylindrical source. This follows from a comparison with the gravitational field of Newtonian cylinders,
or the source can be constructed directly using the cut-and-paste method \cite{Bicak+Zofka_2002}. However, the ranges of the metric parameters where this applies are rather restricted and, beyond these, the analogy fails. In addition to the cosmological constant, both spacetimes are characterized by two independent parameters. The first is associated with the conical defect or missing angle along the axis of symmetry \cite{Vilenkin_1981,Hiscock_1985,Gott_1985}. The second is related to the linear mass density of the source or singularity located along the axis of symmetry. It is the latter what makes the spacetime particularly difficult to interpret. Its different ranges significantly affect the geometric and physical properties of the spacetime, making it challenging to understand, but all the more intriguing.

Can we then, perhaps, find a more intuitive interpretation of the solution based on the behavior of geodesic trajectories? Studying geodesic motion provides a powerful tool for understanding the dynamics of a spacetime, as it is perhaps the closest we can get to an actual experiment. In this paper, following previous works on the topic \cite{Brito2014} and \cite{Brito_2015}, we investigate how different values of the metric parameters affect the existence and nature of geodesics, with an emphasis on motion along the cylindrical coordinates, which coincide with the orbits of the Killing vector fields. The results ultimately lead us to propose a different global interpretation of the solution.

The paper is organized as follows: In \autoref{the_spacetime}, we sum up the properties of the LC$\Lambda$ solution, its symmetries, and singularities. In \autoref{geodesics}, we then analyze the behavior of geodesics, discussing separately the radial, azimuthal, and axial ones, and also the motion of particles dropped from rest in the coordinate system, we study the associated velocities and their dependence on the metric parameters. We then summarize our results and draw conclusions from them in \autoref{conclusions}.

\section{The spacetime}
\label{the_spacetime}
The Linet-Tian or LC$\Lambda$ metric in cylindrical coordinates $t, z, r \in \mathbb{R}, \varphi \in {[}0,2\pi)$ is
\begin{equation}\label{the_metric}
 \begin{array}{rcl}
 {\rm d}s^2 & = & Q(r)^{\frac{2}{3}} \left( -P(r)^{-2\frac{4\sigma^2-8\sigma+1}{3A}} {\rm d}t^2
 + P(r)^{2 \frac{8\sigma^2-4\sigma-1}{3A}} {\rm d}z^2 \right.\\
 && \left. + P(r)^{-4 \frac{2\sigma^2+2\sigma-1}{3A}} \frac{{\rm d}\varphi^2}{C^2} \right) +{\rm d}r^2,
 \end{array}
\end{equation}
with
\begin{equation}\label{Definition_of_A_using_sigma}
A := 4\sigma^2-2\sigma+1
\end{equation}
and
\begin{equation}\label{Metric, positive Lambda}
P(r)= \frac{2 \special_radius}{\pi} \tan \left( \frac{\pi}{2} \frac{r}{\special_radius} \right), \:\: Q(r)= \frac{\special_radius}{\pi} \sin \left( \pi \frac{r}{\special_radius} \right)
\end{equation}
for $\Lambda>0$, while for $\Lambda<0$ we have
\begin{equation}\label{Metric, negative Lambda}
P(r)= \frac{2 \special_radius}{\pi} \tanh \left( \frac{\pi}{2} \frac{r}{\special_radius} \right), \:\: Q(r)= \frac{\special_radius}{\pi} \sinh \left( \pi \frac{r}{\special_radius} \right),
\end{equation}
where we defined the typical length scale
\begin{equation}\label{length_scale}
 \special_radius := \frac{\pi}{\sqrt{3 | \Lambda |}}.
\end{equation}
The parameter $C$ determines the angular deficit around the $z$-axis; by choosing it appropriately, we can remove the associated conical singularity. In the following, we will put $C=1$ for simplicity, since it can always be transformed away locally by redefining the range of the coordinate $\varphi$ or, put differently, any expression involving $\varphi$ must be divided by $C$. In the limit $r \rightarrow 0,$ the metric (\ref{the_metric}) becomes the original LC metric of \cite{Levi-Civita} with $m = 2 \sigma$, so that regardless of the sign of $\Lambda$ both solutions behave the same way near the axis. Likewise, in the limit $\Lambda \rightarrow 0$, the metric (\ref{the_metric}) is the LC metric again. The solution has also been shown \cite{Gleiser_2021} to be a subset of rotating perfect fluid spacetimes by Krasi\'{n}sky \cite{Krasinski_1,Krasinski_2} related to the Kasner solution \cite{Kasner}.
\subsection{Symmetries}\label{symmetries}
The metric components in (\ref{the_metric}) depend solely on the radial coordinate $r$ which implies the existence of three Killing vectors: $\partial t$, $\partial z,$ and $\partial \varphi$. In addition to these, extra Killing vectors arise for special values of $\sigma$. In particular, we have
\begin{eqnarray}
\varphi \partial z - C^2 z\partial \varphi \quad &\mathrm{for} \quad \sigma = \pm \frac{1}{2}, \label{additional_killing_sigma_=+-1/2}\\
z\partial t + t\partial z \quad &\mathrm{for} \quad \sigma = 0, \sigma = 1 \label{additional_killing_sigma=0_1},\\
\varphi \partial t + C^2 t \partial \varphi \quad &\mathrm{for} \quad \sigma = \frac{1}{4} \label{additional_killing_sigma=1/4}.
\end{eqnarray}
There are additional symmetries leaving the metric unchanged: $\sigma \rightarrow 1/4 \sigma$ swaps the roles of $z$ and $\varphi$. Therefore, $\sigma= \pm 1/2$ is rather a planar case than a cylindrical one. This agrees with the relation (\ref{additional_killing_sigma_=+-1/2}) above. Next, $\sigma \rightarrow \sigma / (2\sigma-1)$, interchanges $z$ and $t$, with $\sigma= 0, \sigma= 1$ corresponding thus to $z$-boost symmetric solutions (\ref{additional_killing_sigma=0_1}). Finally, replacing $\sigma \rightarrow 1/2 - \sigma$ yields the roles of $\varphi$ and $t$ interchanged. This indicates that the case $\sigma= 1/4$ corresponds to a $\varphi$-boost symmetric solution, confirming relation (\ref{additional_killing_sigma=1/4}).

Furthermore, with $\Lambda>0$, the solution is reflection-symmetric with respect to any cylindrical surface $r=k \special_radius, k$ integer. In the following and without loss of generality, we will thus restrict the range of $r$ to $[0, \special_radius]$ with $r=0$ equivalent to $r=2k \special_radius$ and $r=\special_radius$ equivalent to $r=(2k+1) \special_radius$, with $k$ integer. Note that the coordinate transformation $r \rightarrow \special_radius - r$ leaves the sines in (\ref{Metric, positive Lambda}) unchanged while turning the tangent into cotangent, which is just the reciprocal value of the tangent. If we then find a $\sigma$ that changes the sign of the exponents in (\ref{the_metric}), the resulting metric is the same as the original. In fact, if we want to keep all terms unchanged, there is only one non-trivial solution $\sigma \rightarrow (1-\sigma)/(1-4\sigma)$ but if we admit switching $z \leftrightarrow \varphi$ we find another option: $\sigma \rightarrow (1-4\sigma)/4(1-\sigma)$. In conclusion, various values of $\sigma$ in (\ref{the_metric}) represent the same spacetime. To argue that the metric has a symmetry interchanging the roles of $z$ and $\varphi$, both coordinates must have the same range of values, which is not the case here. We will return to this point later on, in \autoref{Axial geodesics}.

\subsection{Singularities}
Inspecting the values of the Kretschmann scalar, $K = R_{\mu \nu \kappa \lambda} R^{\mu \nu \kappa \lambda},$ we find singular cylindrical surfaces depending on the sign of $\Lambda$ and the value of $\sigma$. For $\Lambda > 0,$ we get two possible singularities at $r = 0$ and $r = \special_radius$, with $\special_radius$ defined by the relation (\ref{length_scale}) and a finite proper distance $\mathcal{R}$ between the singularities. Inspecting these surfaces, we find that the Kretschmann scalar vanishes at $r = 0$ for $\sigma = 0$ and $\sigma = 1/2$, and that it vanishes at $r = \special_radius$ for $\sigma = -1/2$ and $\sigma = 1/4,$ so that no singularity occurs there for these values of $\sigma$.

For $\Lambda < 0$, the behavior of the Kretschmann scalar is the same near the axis since both spacetimes have the same asymptotic form here, namely the LC one. Therefore, we only need to analyze the limit $r \to \infty$. We find that $K$ remains finite here for all values of $\sigma$ with no singularity at radial infinity. This is not surprising since the spacetime approaches AdS asymptotically \cite{Zofka+Bicak_2008}. These results are included in \autoref{Singularities}. Although the solution is formally periodic for $\Lambda > 0$, the singular cylindrical surfaces prevent an observer from crossing into the adjacent regions, and the solution thus does not extend to radial infinity.
\subsection{Proper lengths}\label{Proper lengths}
\begin{figure}[ht]
\begin{center}
\begin{subfigure}[c]{.8\textwidth}
 \begin{subfigure}{\textwidth}
 \begin{subfigure}{.45\textwidth}
 \includegraphics[width=\linewidth]{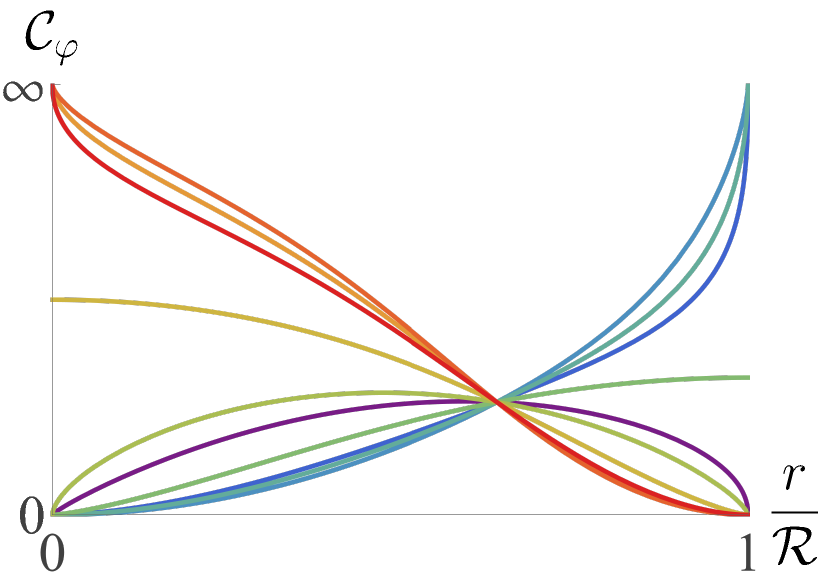}
 \caption{$\Lambda > 0$, azimuthal direction.}
 \end{subfigure}
 \begin{subfigure}{.45\textwidth}
 \includegraphics[width=\linewidth]{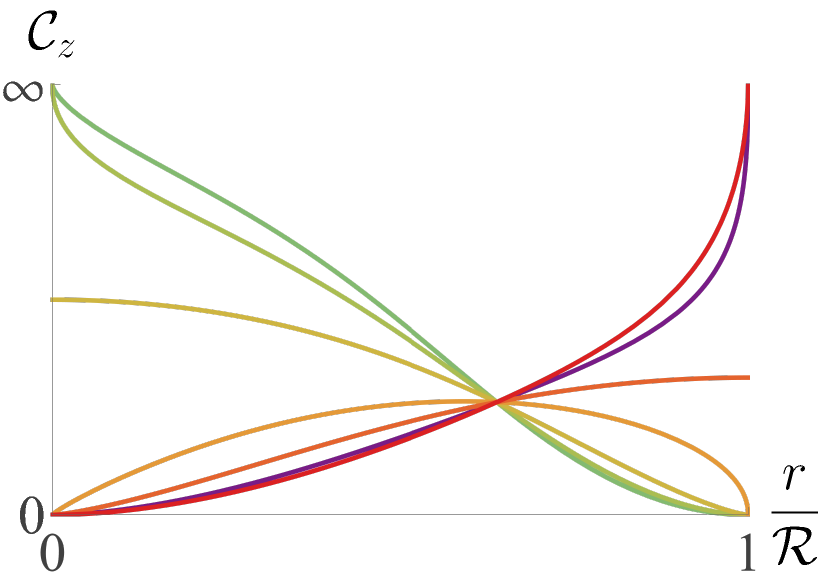}
 \caption{$\Lambda > 0$, axial direction.}
 \end{subfigure}
 \end{subfigure}
 \begin{subfigure}{\textwidth}
 \begin{subfigure}{.45\textwidth}
 \includegraphics[width=\linewidth]{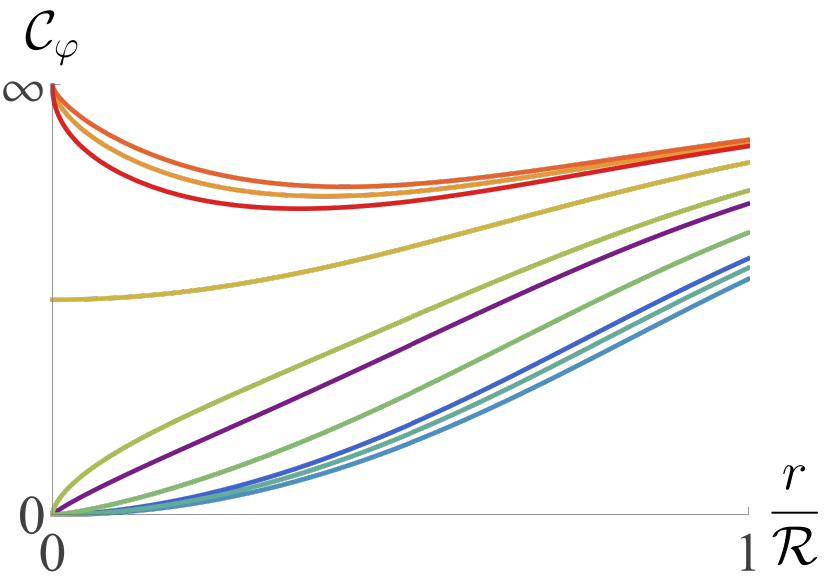}
 \caption{$\Lambda < 0$, azimuthal direction.}
 \end{subfigure}
 \begin{subfigure}{.45\textwidth}
 \includegraphics[width=\linewidth]{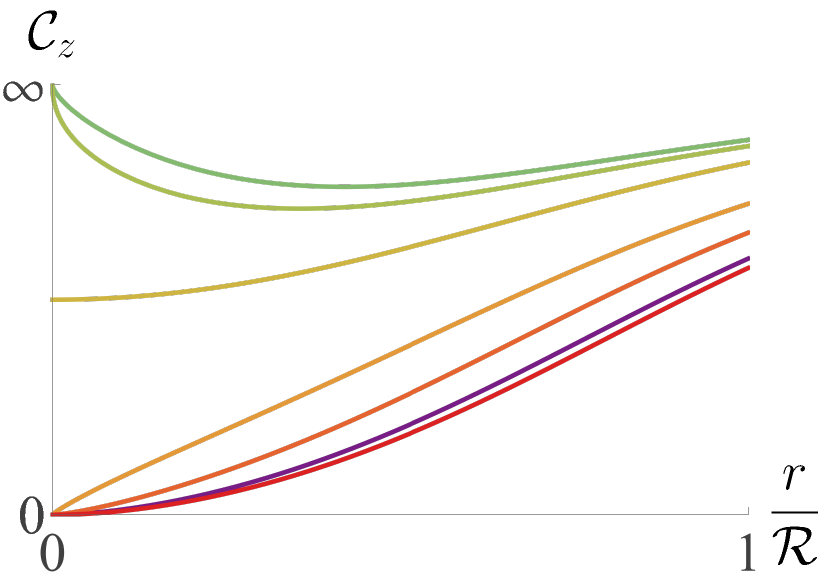}
 \caption{$\Lambda < 0$, axial direction.}
 \end{subfigure}
 \end{subfigure}
\end{subfigure}
\hspace{-1cm}
\begin{subfigure}[c]{.15\textwidth}
 \includegraphics[width=\textwidth]{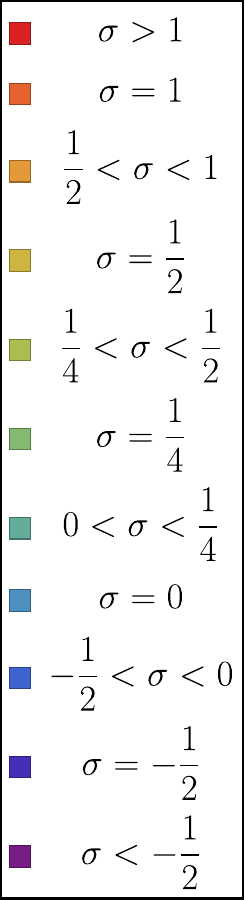}
\end{subfigure}
\end{center}
\caption{\label{proper_lengths}Proper lengths along coordinate axes $\varphi$ and $z$ as functions of the position for various values of $\sigma$. The horizontal axis is the radial coordinate scaled by $\special_radius$ of (\ref{length_scale}). For $\Lambda < 0$, both $\mathcal{C}_\varphi$ of (\ref{Proper circumference, phi}) and $\mathcal{C}_z$ of (\ref{Proper circumference, z}) diverge as $r \rightarrow \infty$. Near $r=0$, all the curves behave as in the LC spacetime \cite{Levi-Civita} regardless of the sign of $\Lambda$. The intervals of $\sigma$ we used are given by the special values $\sigma = -1/2,0,1/4,1/2,1$ that also appear in the relations (\ref{additional_killing_sigma_=+-1/2})-(\ref{additional_killing_sigma=1/4}). Some of the curves overlap---those with a higher value of $\sigma$ define the resulting color. Notice also the symmetry between the azimuthal and axial circumferences.}
\end{figure}

As mentioned above and discussed in \cite{Zofka+Bicak_2008} and \cite{Griffiths_2010}, there exist transformations of $\sigma$ under which the roles of $z$ and $\varphi$ are interchanged. Due to this symmetry and the fact that the metric (\ref{the_metric}) was derived from Einstein equations, which are local differential equations, one could argue that the range of the $z$ coordinate can just as well be finite, say $(0,2\pi]$, the same as for the $\varphi$ coordinate, with the endpoints identified. Based on these considerations, we now define proper circumferences along both the $\varphi$ and $z$ coordinates to further help us analyze and interpret the results of later sections:
\begin{equation}\label{Proper circumference, phi}
\mathcal{C}_\varphi = \int_{0}^{2\pi} \sqrt{g_{\varphi\varphi}} d\varphi = 2 \pi \sqrt{g_{\varphi\varphi}}
\end{equation}
and
\begin{equation}\label{Proper circumference, z}
\mathcal{C}_z = \int_{0}^{2\pi} \sqrt{g_{zz}} dz = 2 \pi \sqrt{g_{zz}}.
\end{equation}
The behavior of these circumferences is shown in \autoref{proper_lengths}. Note that there can be more than just one axis in the spacetime: indeed, the axis is defined as the set of points left invariant by rotations due to the Killing vector $\partial \varphi$. Therefore, if the circumference $\mathcal{C}_\varphi$ vanishes at some $r$, then this particular cylindrical surface is, in fact, an axis. It follows then that for $\Lambda >0$ and $\sigma < -1/2$ or $\sigma \in (1/4,1/2)$ the spacetime has two singular axes with a surface of maximum circumference between them as can be seen in \autoref{proper_lengths}.
{\renewcommand{\arraystretch}{1.3} 
\begin{table}[ht]
\begin{centering}
\begin{tabular}{||c|c|c|c|c|c||}
\hline
\hline
$\Lambda$ & $r$ & $\sigma$ & $\mathcal{C}_\varphi$ & $\mathcal{C}_z$ & Singularity \\
\hline
\hline
\multirow{5}{*}{any $\Lambda$} & \multirow{5}{*}{$r = 0$} & $\sigma < 0$ & $\mathcal{C}_{\varphi}=0$ & $\mathcal{C}_z=0$ & Yes \\
 & & $\sigma = 0$ & $\mathcal{C}_{\varphi}=0$ & $\mathcal{C}_z$ finite & No \\
 & & $\sigma \in (0, \frac{1}{2})$ & $\mathcal{C}_{\varphi}=0$ & $\mathcal{C}_z \to \infty$ & Yes \\
 & & $\sigma = \frac{1}{2}$ & $\mathcal{C}_{\varphi}$ finite & $\mathcal{C}_z$ finite & No \\
 & & $\sigma > \frac{1}{2}$ & $\mathcal{C}_{\varphi} \to \infty$ & $\mathcal{C}_z=0$ & Yes \\ \cline{2-6}
\cline{2-5}
\hline
\hline
\multirow{8}{*}{$\Lambda > 0$}
 & \multirow{8}{*}{$r = \special_radius$} & $\sigma < -\frac{1}{2}$ & $\mathcal{C}_{\varphi}=0$ & $\mathcal{C}_z \to \infty$ & Yes \\
 & & $\sigma = -\frac{1}{2}$ & $\mathcal{C}_{\varphi}$ finite & $\mathcal{C}_z$ finite & No \\
 & & $\sigma \in (-\frac{1}{2}, \frac{1}{4})$ & $\mathcal{C}_{\varphi} \to \infty$ & $\mathcal{C}_z=0$ & Yes \\
 & & $\sigma = \frac{1}{4}$ & $\mathcal{C}_{\varphi}$ finite & $\mathcal{C}_z=0$ & No \\
 & & $\sigma \in (\frac{1}{4}, 1)$ & $\mathcal{C}_{\varphi}=0$ & $\mathcal{C}_z=0$ & Yes \\
 & & $\sigma = 1$ & $\mathcal{C}_{\varphi}=0$ & $\mathcal{C}_z$ finite & No \\
 & & $\sigma > 1$ & $\mathcal{C}_{\varphi}=0$ & $\mathcal{C}_z \to \infty$ & Yes \\
\hline
\hline
{$\Lambda < 0$} & $r \to \infty$ & any $\sigma$ & $\mathcal{C}_{\varphi} \to \infty$ & $\mathcal{C}_z \to \infty$ & No \\
 \hline
 \hline
\end{tabular}
\caption{\label{Singularities} Curvature singularities and proper lengths along the axes for specific values of $r$ and $\sigma$. The behavior at $r = 0$ is the same as for the LC solution \cite{Levi-Civita} with $\Lambda = 0$.}
\label{Circumferences of Rings}
\end{centering}
\end{table}
}
\section{Geodesics}\label{geodesics}
Let us now explore how freely falling particles move around the spacetime. We thus solve the geodesic equations, which have the form
\begin{eqnarray}
&\ddot{t} + \frac{g_{tt}'}{g_{tt}} \dot{t} \dot{r} = 0, \label{t geodesic}\\
&\ddot{z} + \frac{g_{zz}'}{g_{zz}} \dot{z} \dot{r} = 0, \label{z geodesic}\\
&\ddot{\varphi} + \frac{g_{\varphi \varphi}'}{g_{\varphi \varphi}} \dot{\varphi} \dot{r} = 0, \label{phi geodesic}\\
&\ddot{r} - \frac{g_{tt}'}{2} \dot{t}^2 - \frac{g_{zz}'}{2} \dot{z}^2 - \frac{g_{\varphi \varphi}'}{2} \dot{\varphi}^2 = 0, \label{r geodesic}
\end{eqnarray}
where the dot stands for differentiation with respect to the affine parameter, and the prime means derivative with respect to $r$. We do not write the equations in full since they are rather unwieldy. Any of the equations (\ref{t geodesic}) - (\ref{r geodesic}) can be replaced by the 4-velocity normalization $g_{\mu\nu} \dot{x}^{\mu} \dot{x}^{\nu}=\epsilon$, or equivalently
\begin{equation}\label{velocity normalization}
g_{tt}\dot{t}^2 + g_{zz}\dot{z}^2 + g_{\varphi \varphi}\dot{\varphi}^2 + \dot{r}^2 = \epsilon,
\end{equation}
with $\epsilon = -1,0,1$ for timelike, null, and spacelike worldlines, respectively. Due to the symmetries of the spacetime discussed in \autoref{symmetries}, we have 3 integrals of motion
\begin{eqnarray}
\dot{t} &=& - \frac{E}{g_{tt}}, \label{t geodesic, integrated}\\
\dot{z} &=& \frac{P_z}{g_{zz}}, \label{z geodesic, integrated}\\
\dot{\varphi} &=& \frac{L}{g_{\varphi \varphi}}, \label{phi geodesic, integrated}
\end{eqnarray}
with $E$, $P_z,$ and $L$ constant and corresponding to the energy and linear and angular momenta per unit mass of the test particle, respectively. There are additional integrals of motion for specific values of $\sigma$ due to the Killing vectors (\ref{additional_killing_sigma_=+-1/2}), (\ref{additional_killing_sigma=0_1}), and (\ref{additional_killing_sigma=1/4}), placing algebraic constraints on motion in these cases. Note also that only massive particles can remain static due to the signs in (\ref{velocity normalization}). We now proceed by discussing special cases of geodesics parallel to the axes.
\subsection[Radial geodesics]{Radial geodesics ($\dot{\varphi} = 0, \dot{z} = 0$)}\label{section,radial geodesics}
Let us first assume that the test particle only travels in the radial direction with $P_z=L=0$. Substituting the conserved energy, $E$, into the relation (\ref{velocity normalization}), we obtain a separated radial equation
\begin{equation}
 \dot{r}^2 = \epsilon - \frac{E^2}{g_{tt}}.\label{radial geodesic}
\end{equation}
Since both sides of the equation must be non-negative, the RHS defines an effective potential that determines accessible regions. We now study separately the paths of photons and massive particles.
\subsubsection{Null radial geodesics}
These paths have $\epsilon = 0$ and the explicit form of (\ref{radial geodesic}) is
\begin{equation}\label{null radial geodesic}
\dot{r}^2 = E^2 Q(r)^{-2/3} P(r)^{2(4\sigma^2 - 8\sigma + 1)/3 A}.
\end{equation}
Photons thus have no turning points on radial paths.
\subsubsection{Timelike radial geodesics}\label{section: Timelike radial geodesics}
We can now rewrite the equation (\ref{radial geodesic}) as
\begin{equation}\label{timelike radial geodesic}
\dot{r}^2 = \left[E^2 - V(r)\right] Q(r)^{-2/3}P(r)^{2(4\sigma^2 - 8\sigma + 1)/3A},
\end{equation}
where we introduced the effective potential
\begin{equation}\label{radial potential}
V(r) = Q(r)^{\frac{2}{3}}P(r)^{-2(4\sigma^2 - 8\sigma + 1)/3A}.
\end{equation}
To locate the extrema of (\ref{radial potential}), we calculate its derivative starting with $\Lambda > 0$.
\begin{equation}\label{radial potential derivative, positive Lambda}
V'(r) = \frac{2}{3}\frac{Q(r)^{\frac{2}{3}}P(r)^{\frac{2(4\sigma^2 - 8\sigma + 1)}{3A}}}{\sin\left(\pi \frac{r}{\special_radius}\right)} \!\! \left( \cos\left(\pi \frac{r}{\special_radius}\right) - \frac{4\sigma^2 -8\sigma + 1}{A} \! \right) \!.
\end{equation}
Therefore, a massive particle can remain at rest where $\cos\left(\pi \frac{r}{\special_radius}\right) = (4\sigma^2 -8\sigma + 1) / A$. Such a point only exists if $0\leq \sigma \leq 1/4$ or $\sigma \geq 1$. We plot the effective potential in the \autoref{effective potential plots, Lambda > 0}. Arguing along the same lines with $\Lambda < 0$ now, we simply replace the trigonometric functions by their hyperbolic counterparts to obtain the plot shown in the \autoref{effective potential plots, Lambda < 0}. An extremum is now present if $\sigma \leq 0$.
\begin{figure}[ht]
\begin{subfigure}{.5\textwidth}
\centering
\includegraphics[width=\linewidth]{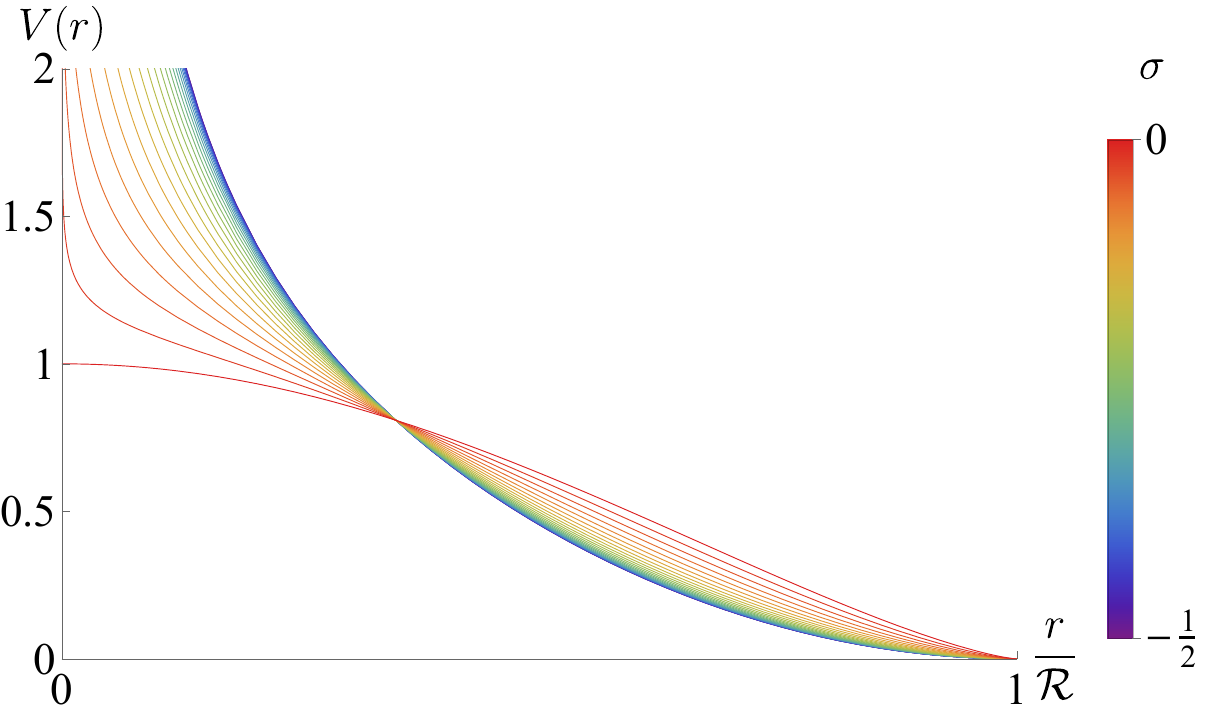}
\caption{$\sigma \leq 0$}\label{effective potential plots, Lambda > 0, sigma < 0}
\end{subfigure}%
\begin{subfigure}{.5\textwidth}
\centering
\includegraphics[width=\linewidth]{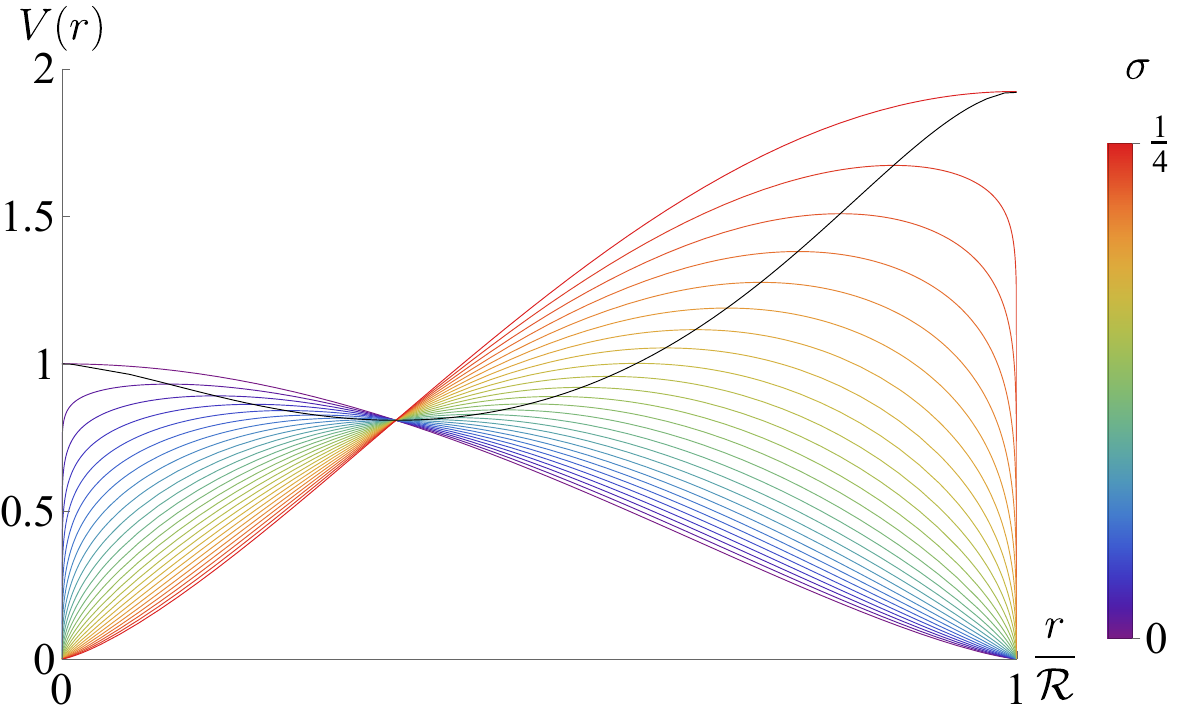}
\caption{$\sigma \in [0, \frac{1}{4}]$}\label{firdfiug 2}
\end{subfigure}
\begin{subfigure}{.5\textwidth}
\includegraphics[width=\linewidth]{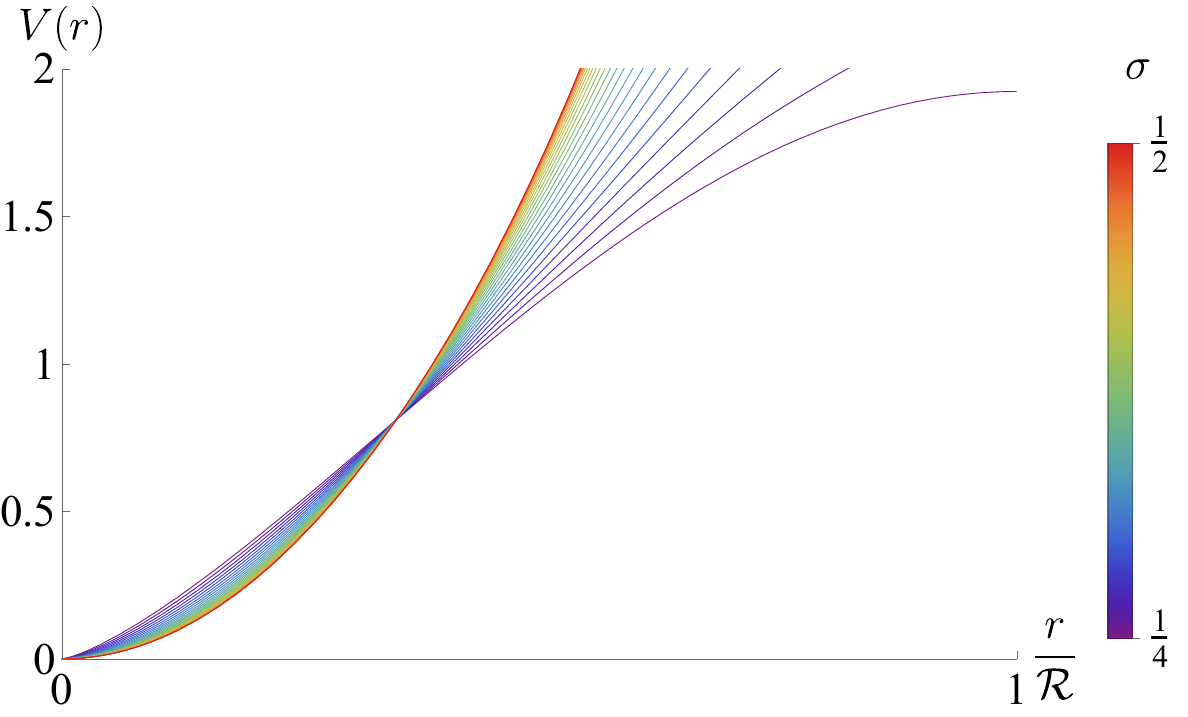}
\caption{$\sigma \in [\frac{1}{4} , \frac{1}{2}]$}\label{firdfiug 3}
\end{subfigure}%
\begin{subfigure}{.5\textwidth}
\includegraphics[width=\linewidth]{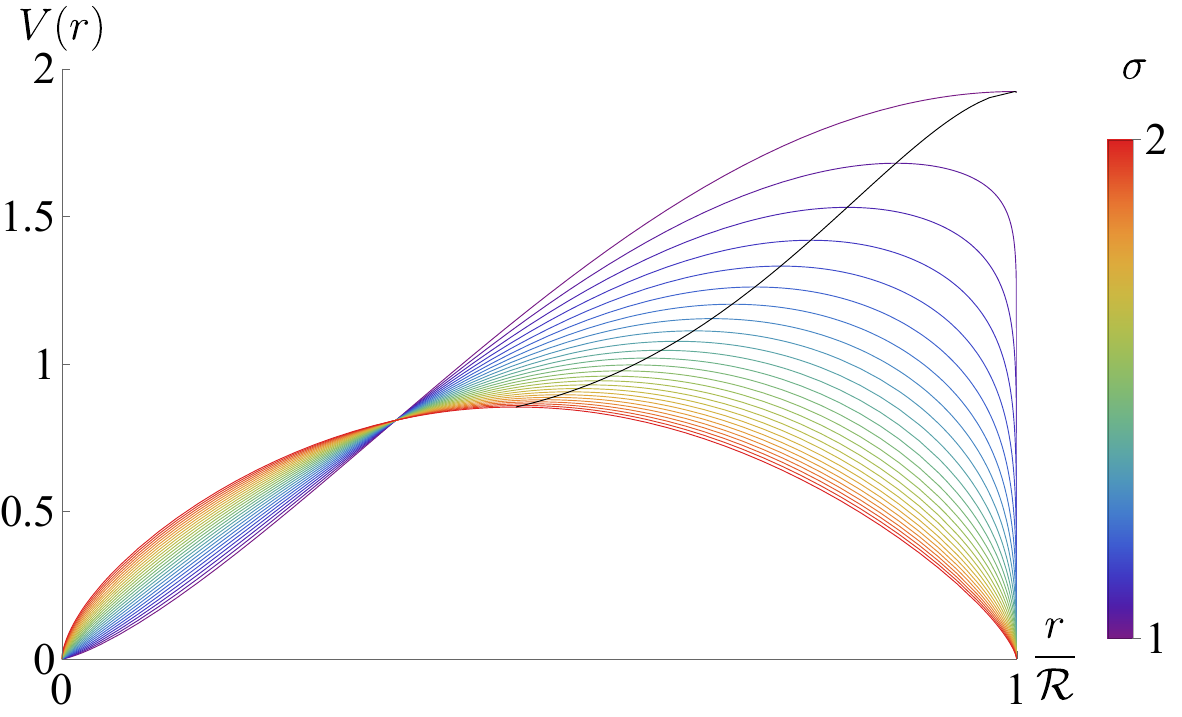}
\caption{$\sigma \geq 1$}\label{firdfiug 4}
\end{subfigure}
\caption{\label{effective potential plots, Lambda > 0}The effective potential (\ref{radial potential}) for purely radial motion with $\Lambda > 0$ (we chose $\Lambda = 0.5$) and various values of $\sigma$, producing the color-coded curves. The black curves highlight the points of extrema (maxima, in fact), if present. (We do not show the range $\sigma \in [\frac{1}{2},1]$ since it displays the same behavior as $\sigma \in [\frac{1}{4},\frac{1}{2}]$ and the curves would overlap.) The common point of all curves is given by the relation $P(r)=1$ and is of no consequence.}
\end{figure}
\begin{figure}[!htb]
 \centering
 \begin{subfigure}{.5\textwidth}
 \centering
 \includegraphics[width=0.9\textwidth]{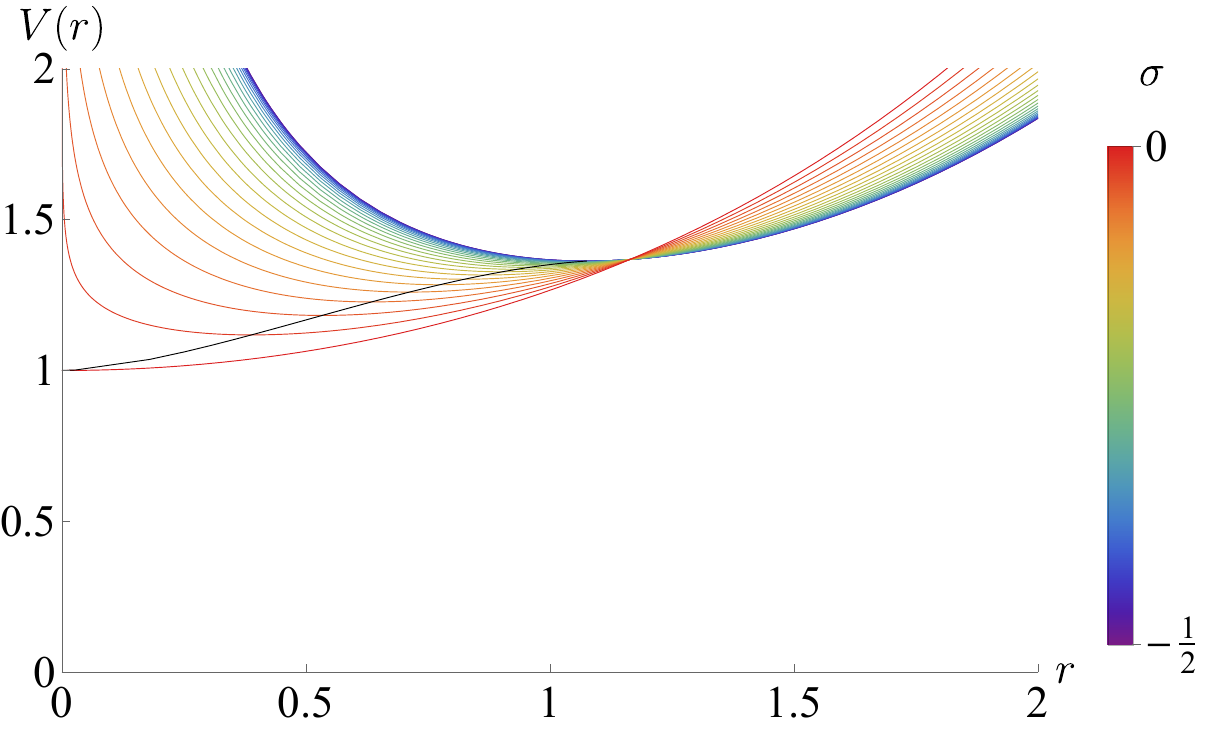}
 \caption{$\sigma \leq 0$}
 \end{subfigure}%
 \begin{subfigure}{0.5\textwidth}
 \centering
 \includegraphics[width=0.9\textwidth]{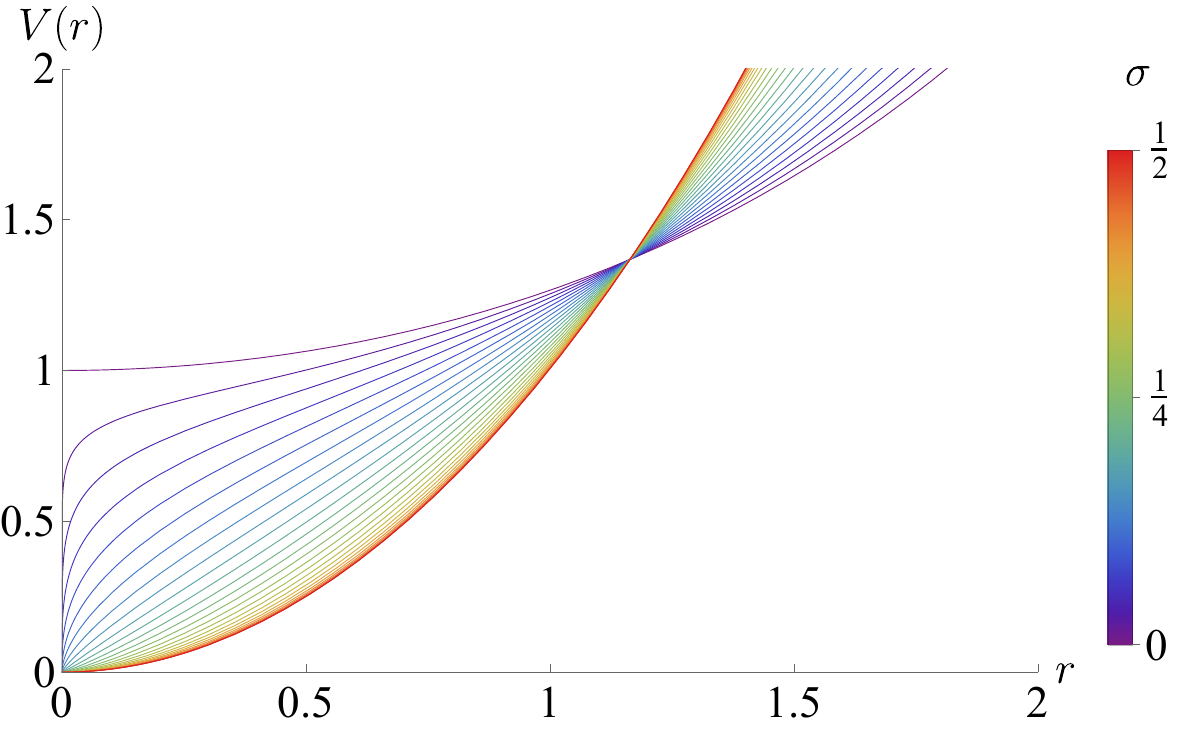}
 \caption{$\sigma \geq 0$}
 \end{subfigure}
 \caption{\label{effective potential plots, Lambda < 0}The effective potential (\ref{radial potential}) for purely radial motion with $\Lambda < 0$ (we chose $\Lambda = -0.5$). The color-coded curves correspond to different ranges of $\sigma$, while the black curve shows the points of extrema (minima), if present.}
\end{figure}

For positive $\Lambda$ and $\sigma < 0$, we can see from \autoref{effective potential plots, Lambda > 0, sigma < 0} that no incoming particle can reach the singularity at $r=0$, indicating its repulsive character. In contrast, the singularity at $r = \special_radius$ is reachable for any value of the energy $E$, suggesting that it is attractive. When $\sigma \in (0,\frac{1}{4})$, the effective potential shows that both singularities are attractive. Due to the maximum in the potential, particles with an energy lower than this maximum will inevitably bounce and fall back toward a singularity. Particles of higher energies can escape and travel to the other singularity. For $\sigma$ from $(\frac{1}{4},1)$, the effective potential resembles the case $\sigma < 0$, but the attractive character of the singularities is now reversed, which means that particles cannot reach the outer singularity at $r = \special_radius$, turning around at some maximum $r$ and falling back toward $r=0$. Finally for $\sigma > 1$, the potential exhibits the same behavior found in the range $0 < \sigma < \frac{1}{4}$, indicating that both singularities are attractive.

In the case of negative $\Lambda$ and $\sigma < 0$, the effective potential has a minimum, trapping the particle between two turning points and suggesting that the singularity along the axis is repulsive. The outer turning is due to the negative cosmological constant. Conversely, for $\sigma > 0$, the potential shows that the singularity is attractive.
\subsubsection{Momentarily static massive particles}
\begin{figure}[!htb]
 \centering
 \begin{subfigure}{.5\textwidth}
 \centering
 \includegraphics[width=0.9\textwidth]{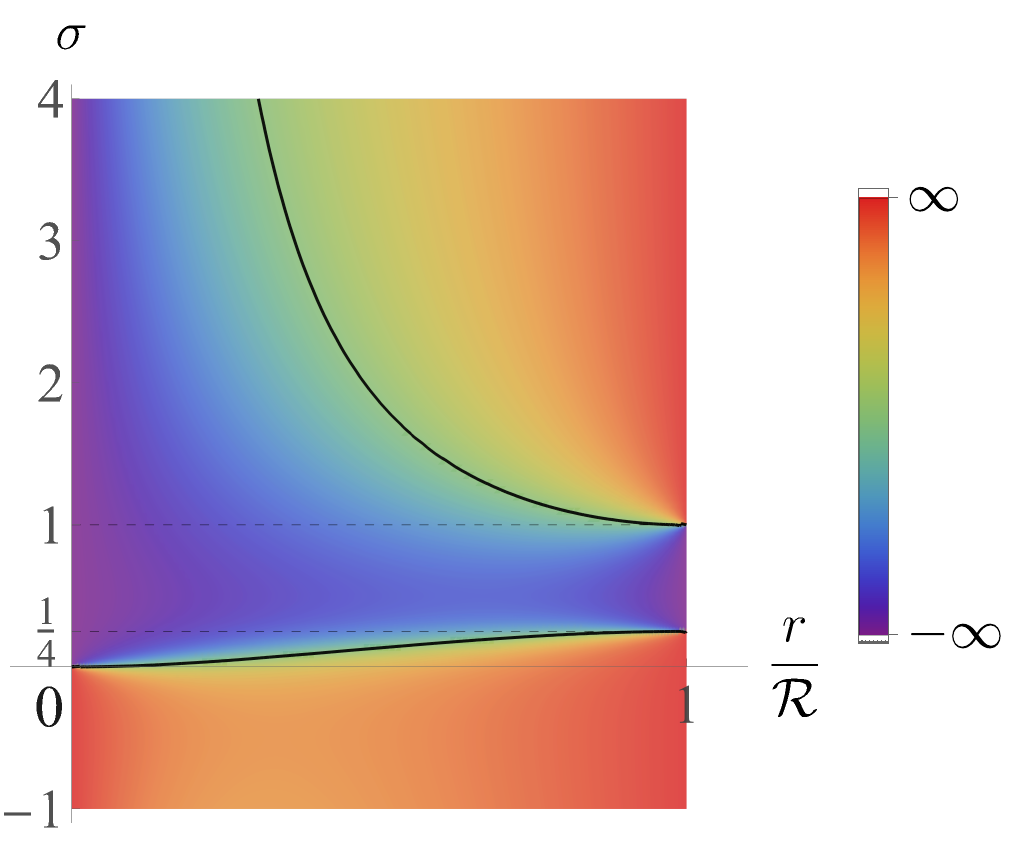}
 \caption{$\Lambda > 0$}\label{Attractive + repulsive, Lambda > 0}
 \end{subfigure}%
 \begin{subfigure}{0.5\textwidth}
 \centering
 \includegraphics[width=0.9\textwidth]{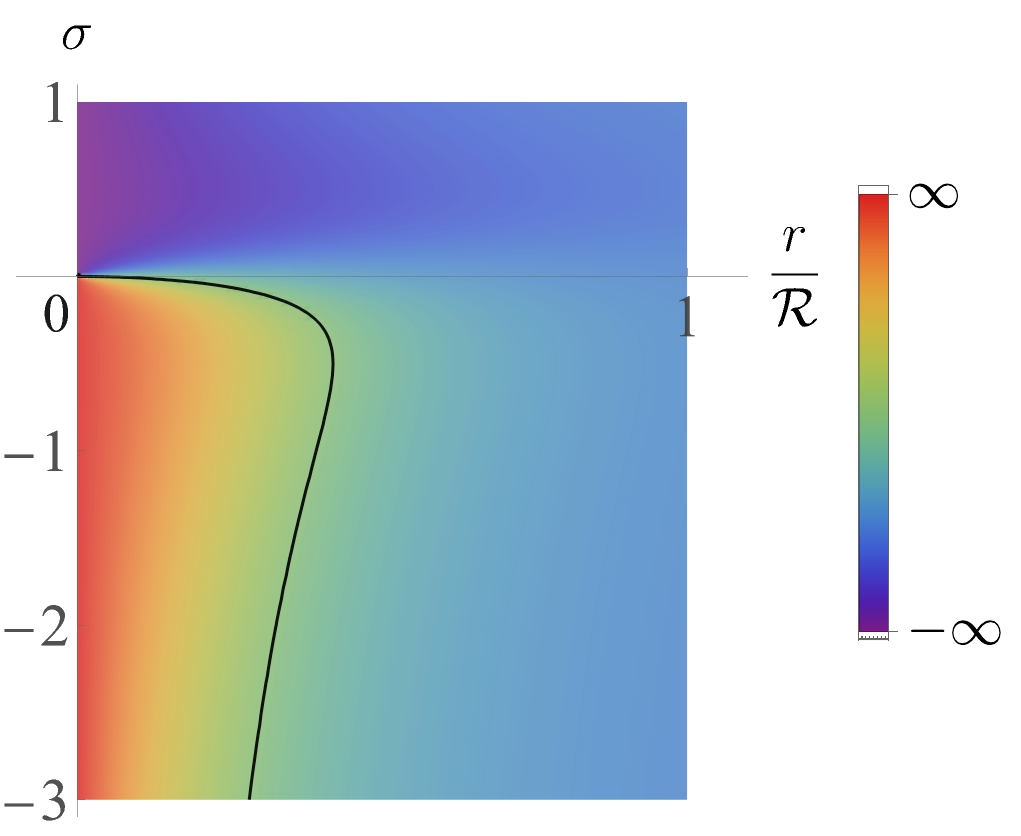}
 \caption{$\Lambda < 0$}\label{Attractive + repulsive, Lambda < 0}
 \end{subfigure}
 \caption{\label{Attractive + repulsive} The radial acceleration of momentarily static particles (\ref{r, acceleration}). Red color means acceleration to the right, while blue acceleration points to the left, green is close to zero. The solid black curves correspond to static particles of zero acceleration and thus also to the black curves denoting the extrema in \autoref{effective potential plots, Lambda > 0} and \autoref{effective potential plots, Lambda < 0}. The horizontal axis is the radial coordinate scaled by $\special_radius$ of (\ref{length_scale}).}
\end{figure}

As a special case of radial motion, we now turn our attention to momentarily static particles that we drop from rest in our coordinate system and calculate in which direction they start moving (necessarily, this must be radial motion due to the relations (\ref{z geodesic, integrated}) and (\ref{phi geodesic, integrated})) to determine whether the singularities are attractive or repulsive. The remaining radial geodesic equation (\ref{r geodesic}) now reduces to
\begin{equation}\label{r, acceleration}
\ddot{r} = - \frac{1}{2} \frac{g'_{tt}}{g_{tt}}.
\end{equation}
For $\Lambda > 0$, we then get
\begin{equation}\label{r, acceleration, Lambda > 0}
\ddot{r} = -\frac{\pi}{3 \special_radius \sin \left(\frac{\pi r}{\special_radius}\right)} \left( \cos\left(\pi \frac{r}{\special_radius}\right) - \frac{4\sigma^2 - 8\sigma + 1}{4\sigma^2 - 2\sigma + 1} \right).
\end{equation}
For $\Lambda < 0$ we simply replace cos with cosh in the above expression. If the sign of the acceleration is positive, the particle is pushed toward higher radii, and vice versa. In \autoref{Attractive + repulsive}, we plot the acceleration as a function of $r$ and $\sigma$ for both signs of the cosmological constant.

Unlike with a zero cosmological constant, here we are dealing with a balance between the interaction with the two singularities and an acceleration due to $\Lambda$. The resulting expression is rather complicated and the sign of the acceleration is not strictly proportional to $\sigma$. However, if we take the limit $\Lambda \rightarrow 0$, this is the case. If we are so close to one of the singularities that the influence of the other singularity and the cosmological constant are negligible, then if the acceleration pulls the test particle toward the singularity in question it makes sense to call the singularity attractive, and to call it repulsive otherwise. We summarize the properties of the singularities in \autoref{Singularity behavior}. Regardless of the sign of $\Lambda$, the singularity at $r=0$ is attractive for $\sigma>0$ and repulsive otherwise. The radial asymptotic region with $\Lambda < 0$ is governed by the cosmological constant as expected, with no singularity. With $\Lambda > 0$ the situation is more nuanced: the outer singularity at $r=\special_radius$ is repulsive for $\sigma \in (1/4,1)$, and attractive otherwise.
{\renewcommand{\arraystretch}{1.3} 
\begin{table}[h]
\begin{centering}
\begin{tabular}{|c|c|c|c|}
\hline
$\Lambda$ & $r$ & $\sigma$ & Sign \\ \hline
\multirow{2}{*}{any $\Lambda$} & \multirow{2}{*}{$r = 0$} & $\sigma < 0$ & repulsive \\
 & & $\sigma > 0$ & attractive \\ \hline
\multirow{2}{*}{$\Lambda > 0$} & \multirow{2}{*}{$r = \special_radius $} & $\sigma < \frac{1}{4}, \sigma > 1$ & attractive \\
 & & $\frac{1}{4} < \sigma < 1$ & repulsive \\ \hline
\end{tabular}
\caption{\label{Singularity behavior}The nature of the singularities: attractive or repulsive?}
\end{centering}
\end{table}
}

To explore how the radial acceleration depends on $\sigma$, we evaluate
\begin{equation}\label{derivative of radial acceleration}
 \frac{\mathrm{d} \ddot{r}}{\mathrm{d} \sigma} = \frac{2 \pi \left(4 \sigma ^2-1\right) \csc \left(\frac{\pi r}{R}\right)}{R \left(4 \sigma ^2-2 \sigma +1\right)^2}.
\end{equation}
Therefore, for $\sigma \in (-1/2,1/2)$, this quantity is negative, it vanishes for $\sigma= \pm 1/2$, and it is positive otherwise. Combining this information with \autoref{Attractive + repulsive, Lambda > 0}, we can see that as $\sigma$ goes from $-\infty$ to $\infty$ the magnitude of the singularity's repulsive effect near $r=0$ increases up to $\sigma = -1/2$, then decreases up to $\sigma=0$ where the singularity switches to an attractive force, which increases up to $\sigma=1/2$ where it starts decreasing again. Analogously, the singularity at $r = \special_radius$ becomes stronger as $\sigma$ goes from $-\infty$ to $-1/2$.

For a similar discussion in the LC case, see Philbin \cite{Philbin}. In general, the in\-flu\-ence of the cosmological constant reveals itself far away from the axis singularity, i.e., for large values of $r$. It is no surprise as for $\Lambda<0$ it was shown by da Silva et al. \cite{da_Silva_et_al} that in this region the metric (\ref{the_metric}) can be transformed into the anti de Sitter metric in horospherical coordinates (see also the Appendix B in \cite{Zofka+Bicak_2008}).
\subsection[Azimuthal geodesics]{Azimuthal geodesics ($\dot{r} = 0, \dot{z} = 0)$}\label{Azimuthal geodesics}
Inserting $P_z = \dot{r} = 0$ into the radial geodesic equation (\ref{r geodesic}) and the 4-velocity normalization (\ref{velocity normalization}), we obtain
\begin{equation}\label{azimuthal geodesic}
\dot{\varphi}^2\left(\frac{g_{tt} g'_{\varphi \varphi}}{g_{tt}'} - g_{\varphi \varphi}\right) + \epsilon = 0.
\end{equation}
\subsubsection{Null azimuthal geodesics}\label{Null azimuthal geodesics}
They are only present if $\sigma = 1/4$, when the bracket in (\ref{azimuthal geodesic}) vanishes. In this case, photons can follow these trajectories at any $r$ and their coordinate angular velocity is $\omega = U^\varphi / U^t = \pm 1$.
\subsubsection{Timelike azimuthal geodesics}\label{Timelike azimuthal geodesics}
\begin{figure}[!htb]
 \centering
 \begin{subfigure}{.5\textwidth}
 \centering
 \includegraphics[width=0.9\textwidth]{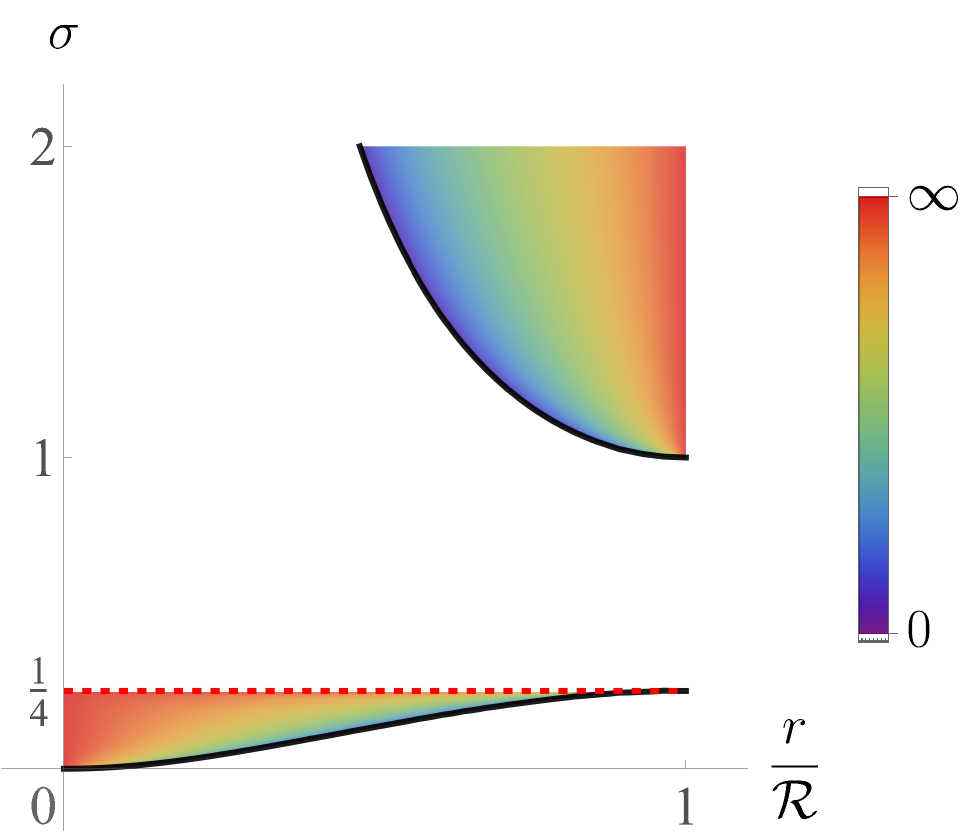}
 \caption{$\Lambda > 0$}\label{azimuthal geodesics Lambda > 0}
 \end{subfigure}%
 \begin{subfigure}{0.5\textwidth}
 \centering
 \includegraphics[width=0.9\textwidth]{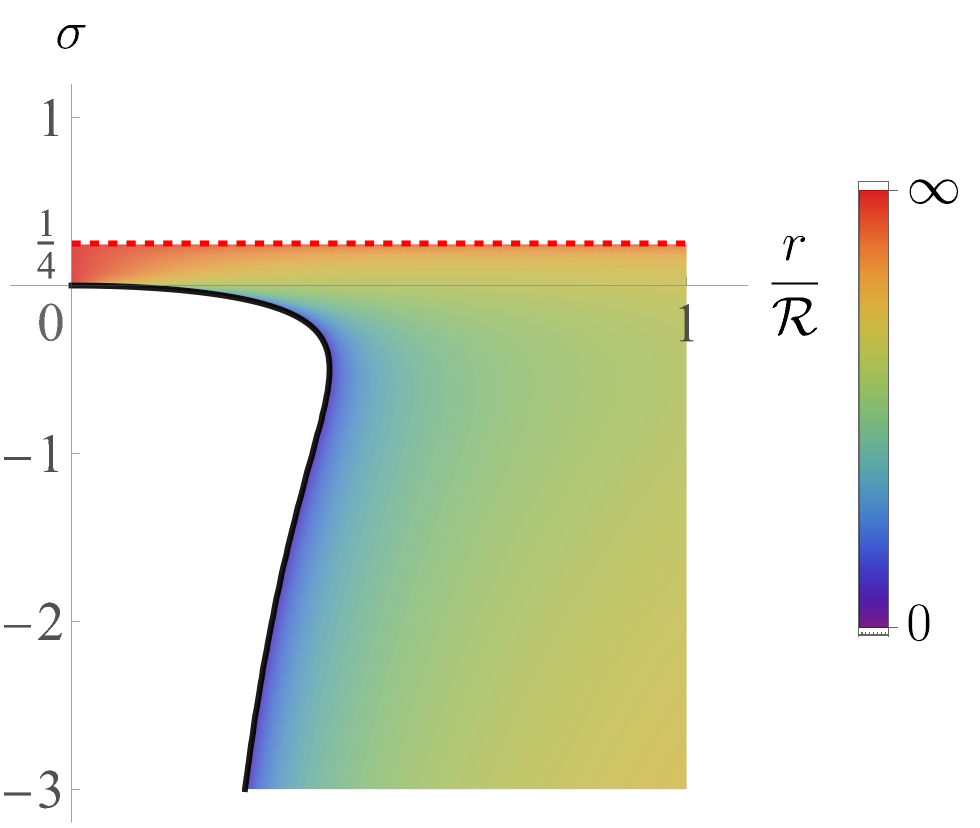}
 \caption{$\Lambda < 0$}\label{azimuthal geodesics Lambda < 0}
 \end{subfigure}
 \caption{\label{azimuthal_geodesics} Angular velocity $\dot{\varphi}(r)$ of azimuthal geodesics (\ref{timelike azimuthal geodesic}) showing where these trajectories exist. The blue color means low velocities, the red color means high velocities. The solid black curves correspond to static particles of zero velocity and coincide with those of (\autoref{Attractive + repulsive}) while the dashed red lines denote null geodesics. The horizontal axis is the radial coordinate scaled by $\special_radius$ of (\ref{length_scale}).}
\end{figure}

In the case of timelike azimuthal geodesics we have $\epsilon = -1$ so that the equation (\ref{azimuthal geodesic}) can be rewritten as
\begin{equation}\label{timelike azimuthal geodesic}
\dot{\varphi}^2 = \frac{g'_{tt}}{g_{tt} g'_{\varphi \varphi} - g_{\varphi \varphi} g'_{tt}}.
\end{equation}
Since $\dot{\varphi}^2 \geq 0$ we get two algebraic conditions on $r$ and $\sigma$ possibly admitting these paths. For $\Lambda > 0$, they are
\begin{eqnarray}
\cos\left(\pi \frac{r}{\special_radius}\right) &\geq \frac{4\sigma^2 - 8\sigma + 1}{4\sigma^2 - 2\sigma + 1}\;\;\mathrm{,}\;\sigma < \frac{1}{4} \label{timelike azimuthal condition, L > 0, sigma < 1/4},\\
\cos\left(\pi \frac{r}{\special_radius}\right) &\leq \frac{4\sigma^2 - 8\sigma + 1}{4\sigma^2 - 2\sigma + 1}\;\;\mathrm{,}\;
\sigma > \frac{1}{4}. \label{timelike azimuthal condition, L > 0, sigma > 1/4}
\end{eqnarray}
However, (\ref{timelike azimuthal condition, L > 0, sigma < 1/4}) can only be satisfied for $0 \leq \sigma < 1/4$, while (\ref{timelike azimuthal condition, L > 0, sigma > 1/4}) can hold only if $\sigma \geq 1$. For $\Lambda < 0$, we get exactly the same relations only with cos replaced by cosh. In this case, the analog of (\ref{timelike azimuthal condition, L > 0, sigma > 1/4}) can never hold as the right-hand side is always less than one for $\sigma > 1/4$. We summarize these results in \autoref{azimuthal_geodesics}.

\begin{figure}[ht]
\begin{subfigure}{.5\textwidth}
\centering
\includegraphics[width=\linewidth]{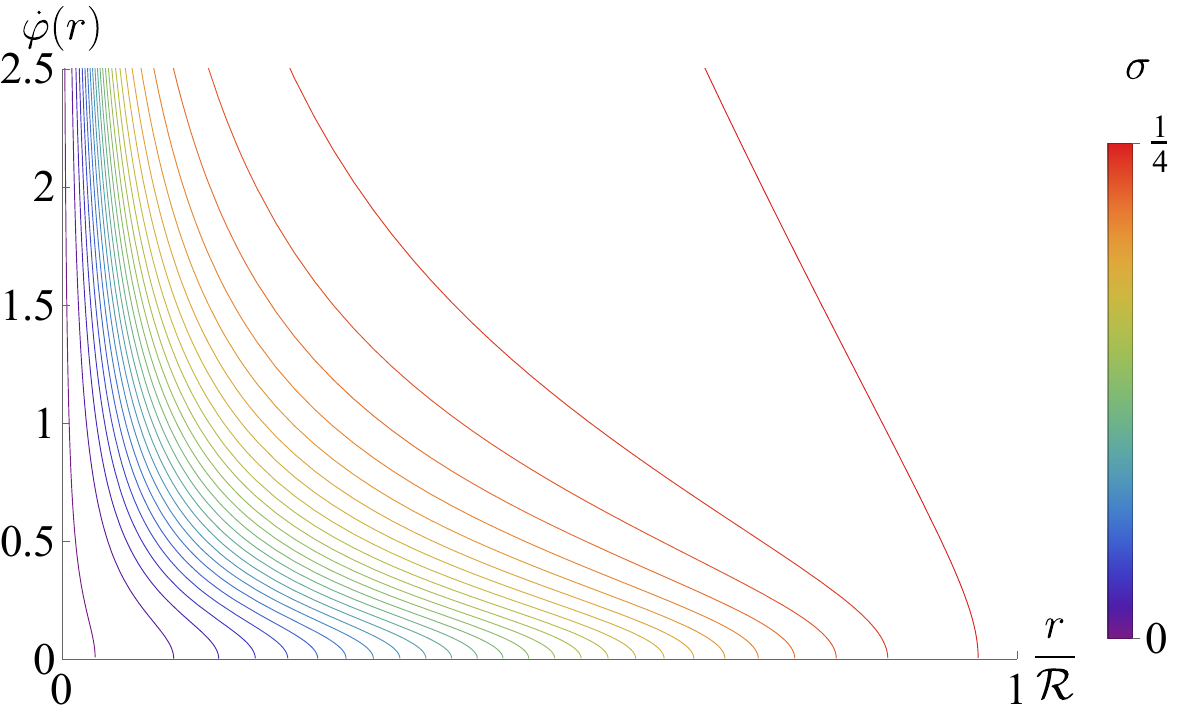}
\caption{{$0 < \sigma < \frac{1}{4}$}}\label{angular velocity plots, Lambda > 0, 0 < sigma < 1/4}
\end{subfigure}%
\begin{subfigure}{.5\textwidth}
\centering
\includegraphics[width=\linewidth]{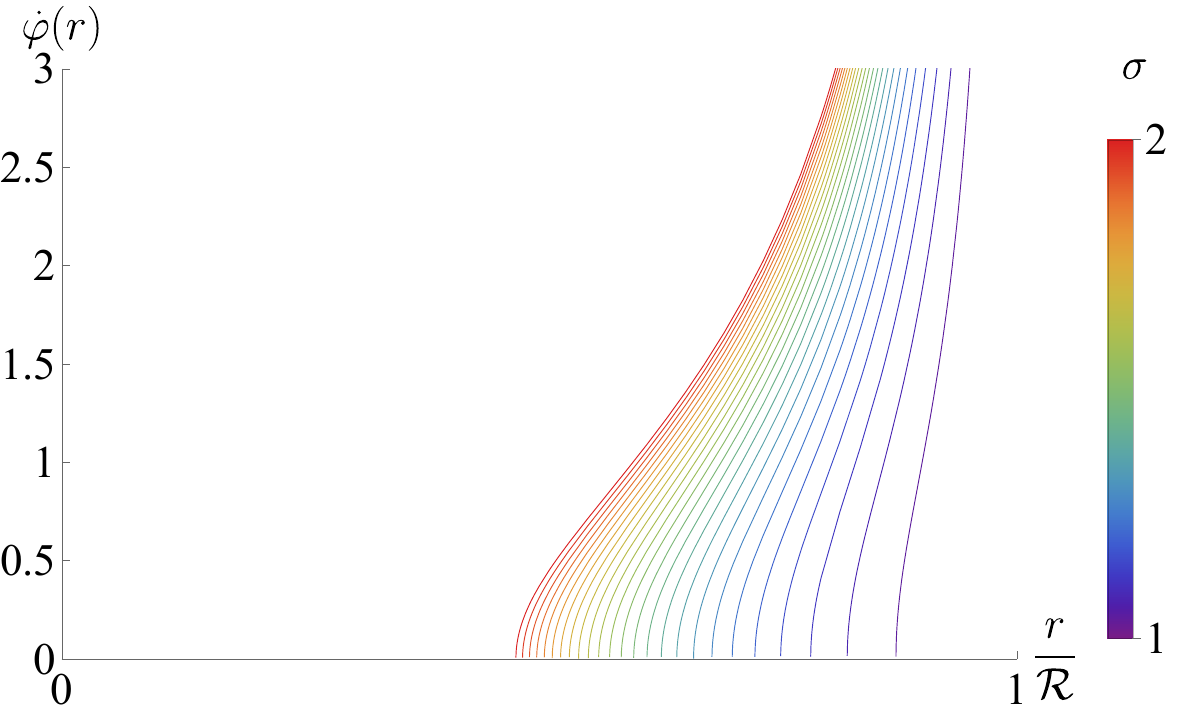}
\caption{$\sigma > 1$}\label{angular velocity plots, Lambda > 0, 1 < sigma}
\end{subfigure}
\caption{\label{angular velocity plots, Lambda > 0}The angular velocity $\dot{\varphi}(r)$ of (\ref{azimuthal geodesic}) for massive particles on azimuthal geodesics with $\Lambda > 0$ (we chose $\Lambda = 0.1$). The curves correspond to various values of $\sigma$ and thus to horizontal cross-sections of \autoref{azimuthal geodesics Lambda > 0}, illustrating the possible values of $\dot{\varphi}(r)$, the corresponding ranges of $r$, and how these depend on $\sigma$. The bottom endpoints of the curves correspond to the black curves of static particles in \autoref{azimuthal geodesics Lambda > 0}.}
\end{figure}
\begin{figure}[ht]
\begin{subfigure}{.5\textwidth}
\centering
\includegraphics[width=\linewidth]{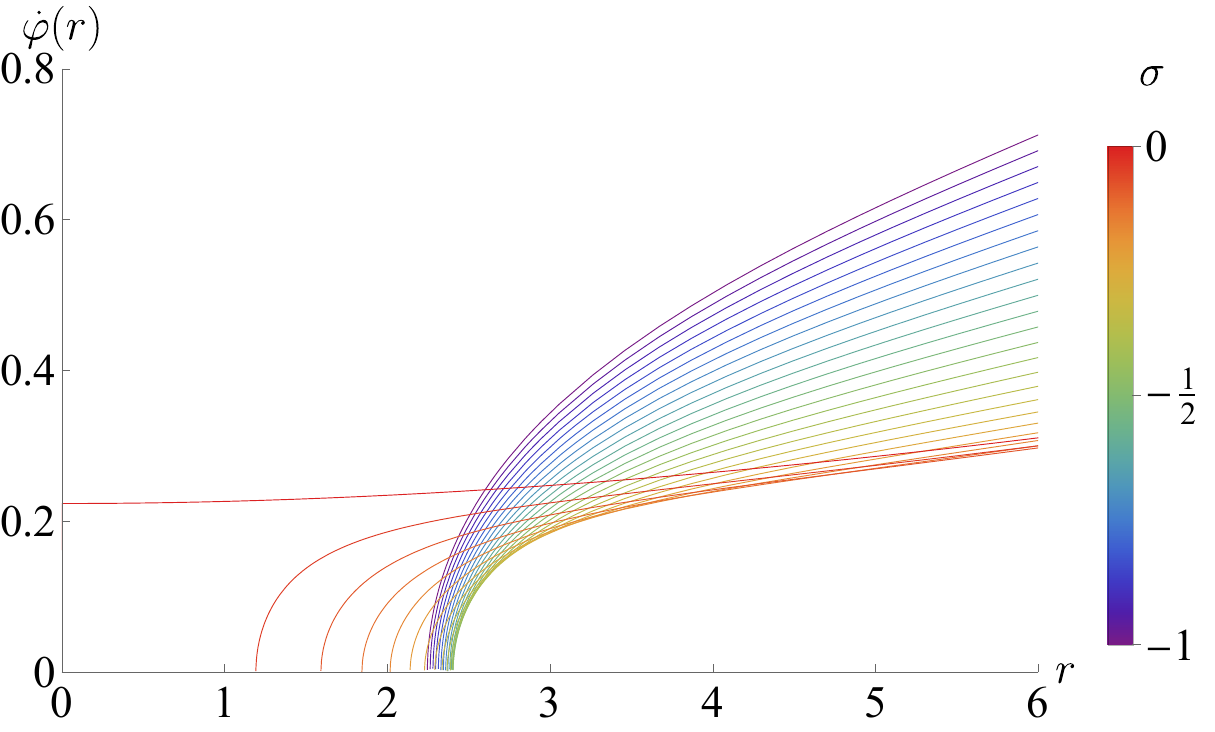}
\caption{$\sigma \leq 0$}\label{angular velocity plots, Lambda < 0, sigma < 0}
\end{subfigure}%
\begin{subfigure}{.5\textwidth}
\centering
\includegraphics[width=\linewidth]{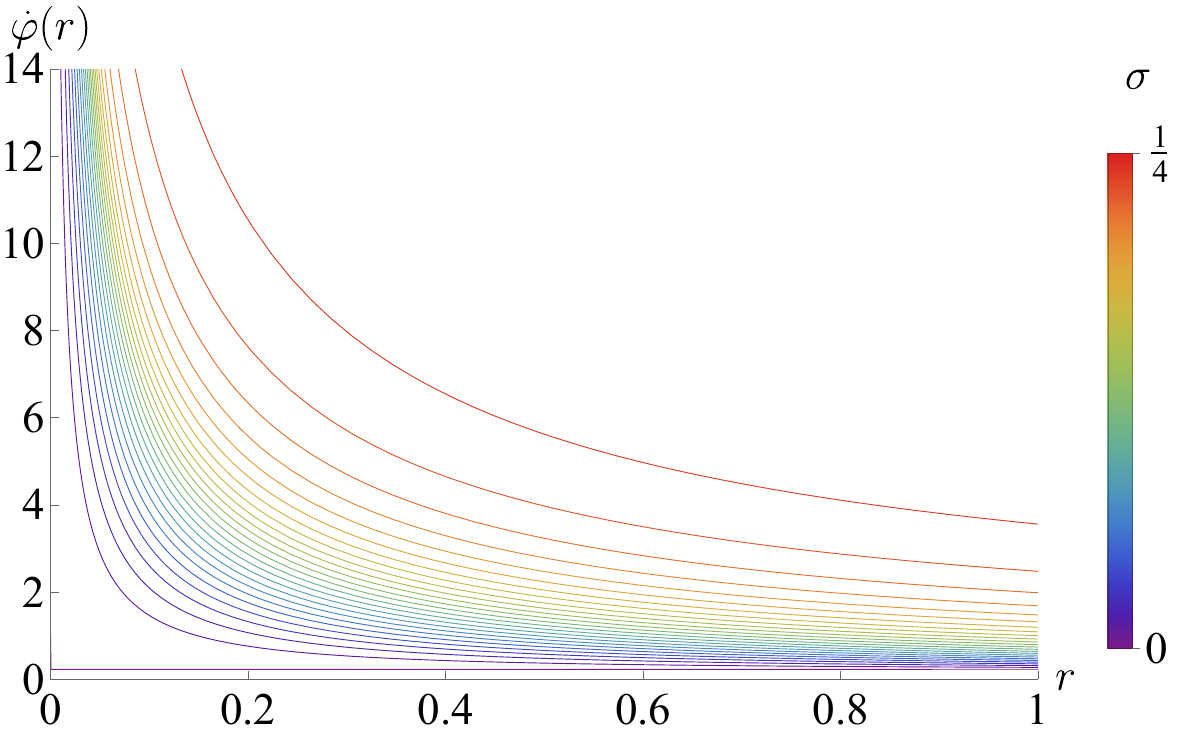}
\caption{$0 \leq \sigma < \frac{1}{4}$}\label{angular velocity plots, Lambda < 0, 0 < sigma < 1/4}
\end{subfigure}
\caption{\label{angular velocity plots, Lambda < 0}The angular velocity $\dot{\varphi}(r)$ of (\ref{azimuthal geodesic}) for massive particles on azimuthal geodesics with $\Lambda < 0$ (we chose $\Lambda = -0.1$). The curves correspond to various values of $\sigma$ and thus to horizontal cross-sections of \autoref{azimuthal geodesics Lambda < 0}, illustrating the possible values of $\dot{\varphi}(r)$, the corresponding ranges of $r$, and how these depend on $\sigma$. The bottom endpoints of the curves in \autoref{angular velocity plots, Lambda < 0, sigma < 0} correspond to the black curves of static particles in \autoref{azimuthal geodesics Lambda < 0}. The two special curves ending on the vertical axis at $\sqrt{-\Lambda/2}$ both correspond to the case $\sigma = 0$.}
\end{figure}
Let us now turn our attention to the angular velocity $\dot{\varphi}$ of these orbits as a function of the coordinate radius $r$, which is plotted for different values of $\sigma$ in \autoref{angular velocity plots, Lambda > 0} and \autoref{angular velocity plots, Lambda < 0}. The bottom endpoints of the colored curves with $\dot{\varphi} = 0$ correspond to static particles and thus also to the solid curves in \autoref{Attractive + repulsive}. Going along these circular paths requires a tug that would counteract the influence of the path's external curvature, which is directed toward larger circumferential radii $\mathcal{C}_\varphi$ and decreases with increasing $\mathcal{C}_\varphi$ (we shall call it the ``centrifugal force''). Whether there can or cannot be a geodesic depends on the outcome of this tug of war. The interpretation of \autoref{angular velocity plots, Lambda > 0, 0 < sigma < 1/4} is clear now: for $\sigma \in [0;1/4]$, $\mathcal{C}_\varphi$ grows both as a function of $r$ and as a function of $\sigma$ so that the closer we get to the attractive singularity at $r=0$ or the larger $\sigma$, the faster we need to run to achieve a balance. This is consistent with the singularity acting like a massive infinite string and no string present for $\sigma = 0$. Perhaps, a similar situation would be expected near the other attractive singularity at $r=\special_radius$, but this is not the case since the centrifugal force points toward higher $r$'s (i.e., toward the singularity) so that balance cannot be achieved. Note also that the attraction of the singularity at $r=0$ increases with increasing $\sigma$ until the geodesic requires the velocity of light and, for $1/4<\sigma<1/2$, the singularity is so strong that it does not admit azimuthal geodesics. This complies with the considerations regarding the change of radial acceleration of static particles (\ref{derivative of radial acceleration}), which gets stronger up to $\sigma=1/2$. The situation is very similar in the range $\sigma > 1$, with the roles of $r$ and $\special_radius$ switched. The circumference $\mathcal{C}_\varphi$ is a decreasing function of $r$ here so that the centrifugal force points toward lower values of $r$ and gets stronger as we approach $r = \special_radius$ (which is now an axis since the circumference $\mathcal{C}_\varphi$ vanishes here) so that there can be no balance near $r=0$. The singular axis at $r = \special_radius$ again acts like a massive string and becomes stronger with increasing $\sigma$, allowing the existence of geodesics. If $\sigma = 1$ then there is no string at $r=\special_radius$, analogously to $\sigma = 0$ and $r=0$. In the ranges $\sigma \in (-1/2;0)$ and $\sigma \in (1/2;1)$, the directions of the centrifugal force and of the force due to the string are the same everywhere so that there can be no balance at any radius. Finally, for $\sigma<-1/2$, balance could, in principle, be achieved at $r= \special_radius$ but as we shall see later, the attraction of the singularity there is too strong to allow azimuthal geodesics.

For $\Lambda < 0$ the situation near $r=0$ is exactly the same as for $\Lambda>0$ since both spacetimes behave like the Levi-Civita solution here while farther away from the axis, geodesics are dominated by the cosmological constant. This extends the region of balance for $\sigma \in [0;1/4)$ up to $r \rightarrow \infty$ and opens up a new region for $\sigma<0$ as seen in \autoref{azimuthal geodesics Lambda < 0}. With $\sigma < 1/4$, the circumference $\mathcal{C}_\varphi$ is an increasing function of $r$ with the centrifugal force pointing toward higher values of $r$. Therefore, for $\sigma \in (0,1/4)$ there will be azimuthal geodesics around the attractive axis at $r=0$, which thus again acts as a massive string, that gets stronger with increasing $\sigma$, see \autoref{angular velocity plots, Lambda < 0, 0 < sigma < 1/4}. For $\sigma < 0$ the situation is different: The singularity is repulsive, trying to prevent the azimuthal geodesics and what now balances the centrifugal force is the cosmological constant $\Lambda$. The singularity's ``cylinder of influence'' starts with a zero radius for $\sigma = 0$, which then increases to a maximum as $\sigma = -1/2$ and then starts shrinking again as $\sigma \rightarrow - \infty$---see \autoref{Attractive + repulsive}. This results in the pattern shown in \autoref{angular velocity plots, Lambda < 0, sigma < 0} with geodesics starting on the axis, receding from it as $\sigma$ decreases, and reaching a maximum distance to ultimately start moving back toward the axis afterward.
\subsection[Axial geodesics]{Axial geodesics ($\dot{r} = 0, \dot{\varphi} = 0)$}\label{Axial geodesics}
It is quite astonishing that the LC$\Lambda$ spacetime admits particles falling freely along the $z$ axis. This is certainly not the case in the Newtonian gravitational field of an infinite massive string, which is usually understood as the Newtonian limit of the LC and LC$\Lambda$ solutions near the axis for $0 < \sigma \ll 1$. Inserting $L = \dot{r} = 0$ into the radial geodesic equation (\ref{r geodesic}) and the 4-velocity normalization (\ref{velocity normalization}), we obtain
\begin{equation}\label{axial geodesic}
\dot{z}^2\left(\frac{g_{tt} g'_{zz}}{g_{tt}'} - g_{zz}\right) + \epsilon = 0.
\end{equation}
\subsubsection{Null axial geodesics}\label{Null axial geodesics}
They are only present if $\sigma = 0$ and $\sigma = 1$, when the bracket in (\ref{axial geodesic}) vanishes. In this case, photons can follow these trajectories at any $r$ and their coordinate axial velocity is $v_z = U^z / U^t = \pm 1$.
\subsubsection{Timelike axial geodesics}\label{Timelike axial geodesics}
\begin{figure}[!htb]
 \centering
 \begin{subfigure}{.5\textwidth}
 \centering
 \includegraphics[width=0.9\textwidth]{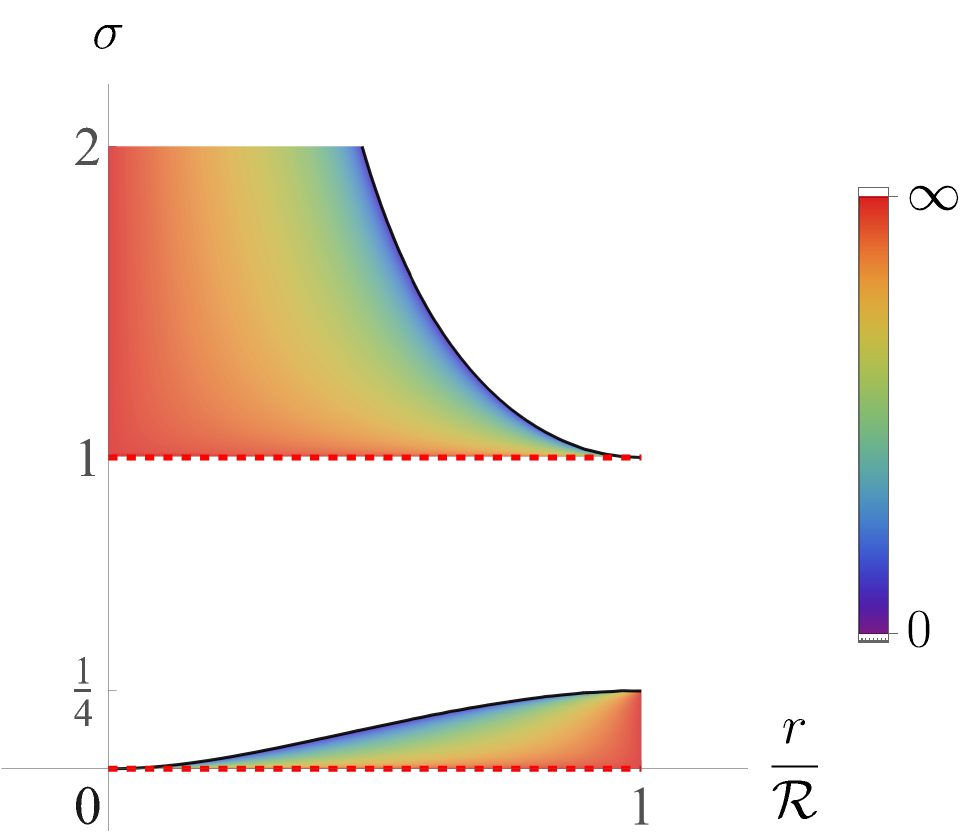}
 \caption{$\Lambda > 0$}\label{axial geodesics Lambda > 0}
 \end{subfigure}%
 \begin{subfigure}{0.5\textwidth}
 \centering
 \includegraphics[width=0.9\textwidth]{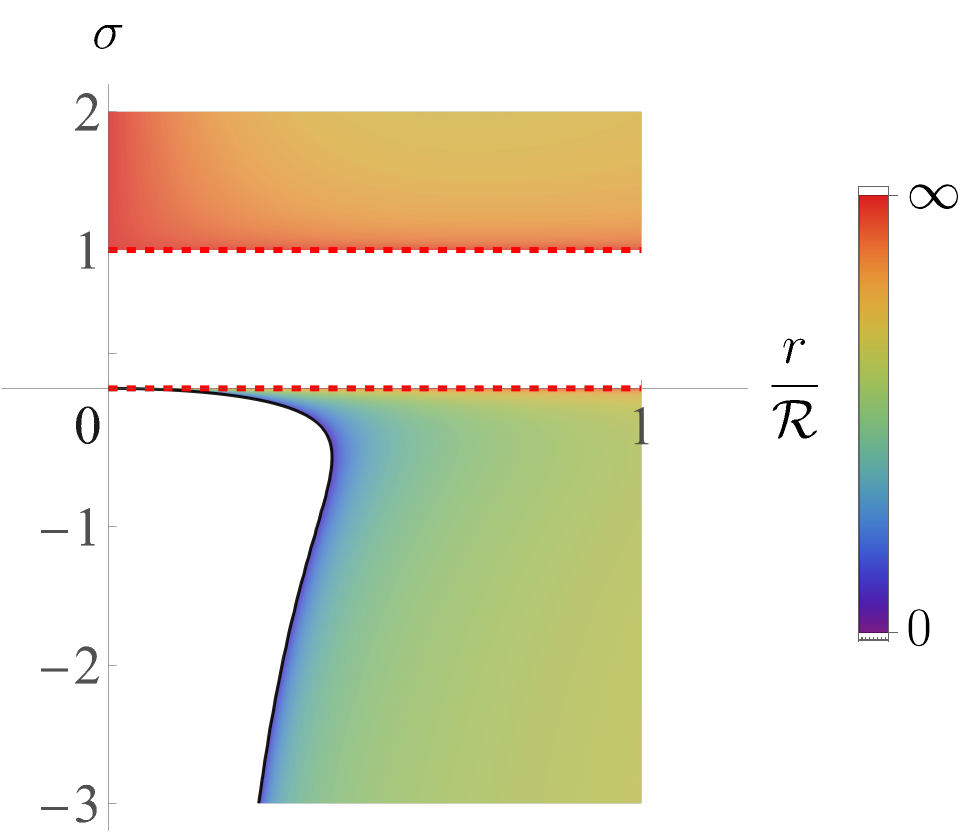}
 \caption{$\Lambda < 0$}\label{axial geodesics Lambda < 0}
 \end{subfigure}
 \caption{\label{axial_geodesics} Axial velocity $\dot{z}(r)$ of axial geodesics showing where these trajectories exist. The blue color means low velocities, the red color means high velocities. The solid black curves correspond to static particles of zero velocity and coincide with those of \autoref{Attractive + repulsive} while the dashed red lines denote null geodesics. The horizontal axis is the radial coordinate scaled by $\special_radius$ of (\ref{length_scale}).}
\end{figure}

Following the same line of reasoning as for the azimuthal motion of \autoref{Timelike azimuthal geodesics}, we obtain the same two conditions (\ref{timelike azimuthal condition, L > 0, sigma < 1/4}) and (\ref{timelike azimuthal condition, L > 0, sigma > 1/4}) but with a different range of $\sigma$. For $\Lambda > 0$, we get
\begin{eqnarray}
\cos\left(\pi \frac{r}{\special_radius}\right) &\geq \frac{4\sigma^2 - 8\sigma + 1}{4\sigma^2 - 2\sigma + 1}\;\;\mathrm{,}\; \sigma \in (-\infty, 0) \cup (1, \infty) \label{timelike axial condition, L > 0, 0 > sigma > 1}, \\
\cos\left(\pi \frac{r}{\special_radius}\right) &\leq \frac{4\sigma^2 - 8\sigma + 1}{4\sigma^2 - 2\sigma + 1}\;\;\mathrm{,}\; \sigma \in (0,1). \label{timelike axial condition, L > 0, 0 < sigma < 1}
\end{eqnarray}
For the inequality (\ref{timelike axial condition, L > 0, 0 > sigma > 1}) to be satisfied, $\sigma$ must be further constrained to $\sigma > 1$. Similarly, the condition (\ref{timelike axial condition, L > 0, 0 < sigma < 1}) holds only for $0 < \sigma \leq 1/4$. For $\Lambda < 0$, the conditions are the same but cos is replaced by cosh. Once again, the inequality (\ref{timelike axial condition, L > 0, 0 < sigma < 1}) can never be satisfied, as the right-hand side is always smaller than the left-hand side within the given range of $\sigma$. We again summarize these results in \autoref{axial_geodesics}.

\begin{figure}[ht]
\begin{subfigure}{.5\textwidth}
\centering
\includegraphics[width=\linewidth]{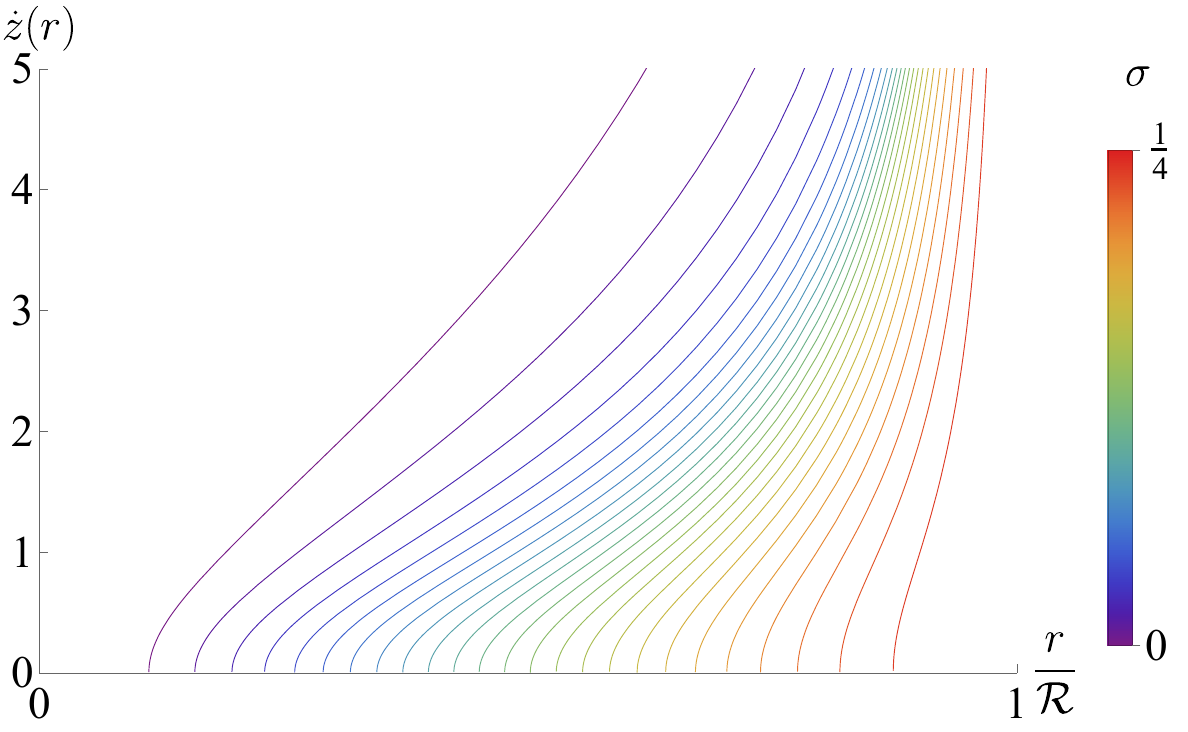}
\caption{$0 < \sigma < \frac{1}{4}$}\label{axial velocity plots, Lambda > 0, 0 < sigma < 1/4}
\end{subfigure}%
\begin{subfigure}{.5\textwidth}
\centering
\includegraphics[width=\linewidth]{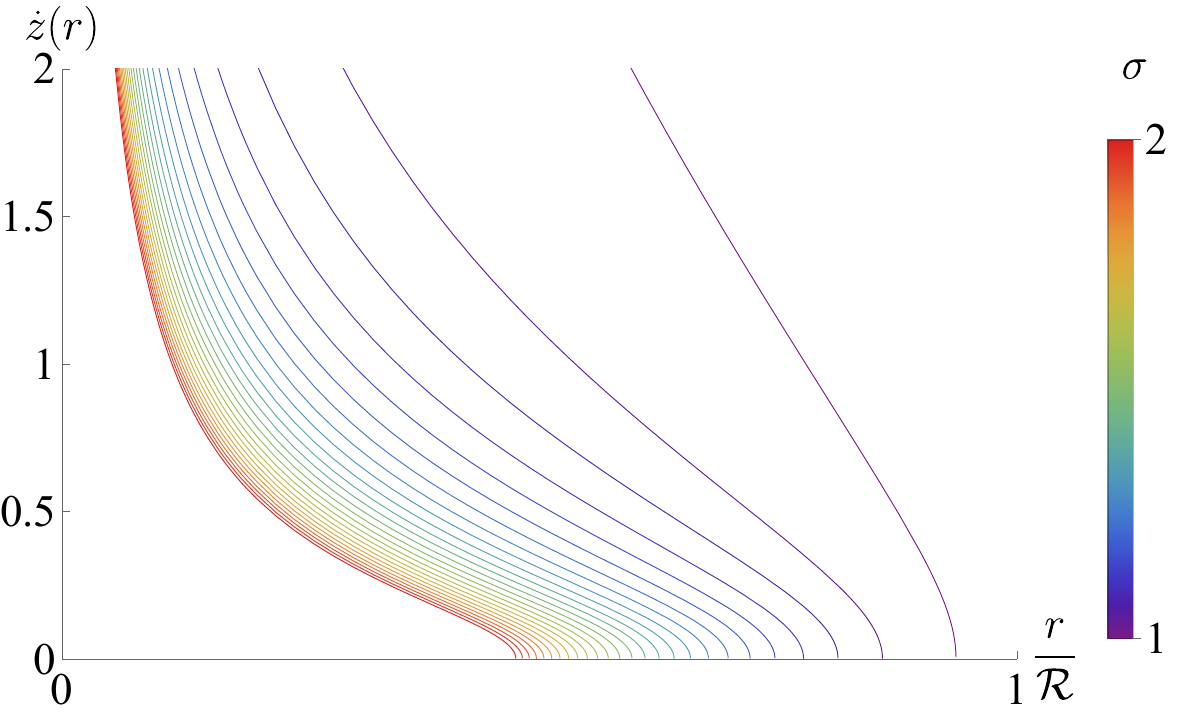}
\caption{$\sigma > 1$}\label{axial velocity plots, Lambda > 0, sigma > 1}
\end{subfigure}
\caption{\label{axial velocity plots, Lambda > 0}The axial velocity for massive particles (\ref{axial geodesic}) with $\Lambda > 0$ (we chose $\Lambda = 0.1$). The curves correspond to various values of $\sigma$ and thus to horizontal cross-sections of \autoref{axial geodesics Lambda > 0}, illustrating the possible values of $\dot{z}(r)$, the corresponding ranges of $r$, and how these depend on $\sigma$. The bottom endpoints of the curves correspond to the black curves of static particles in \autoref{axial geodesics Lambda > 0}.}
\end{figure}
\begin{figure}[ht]
\begin{subfigure}{.5\textwidth}
\centering
\includegraphics[width=\linewidth]{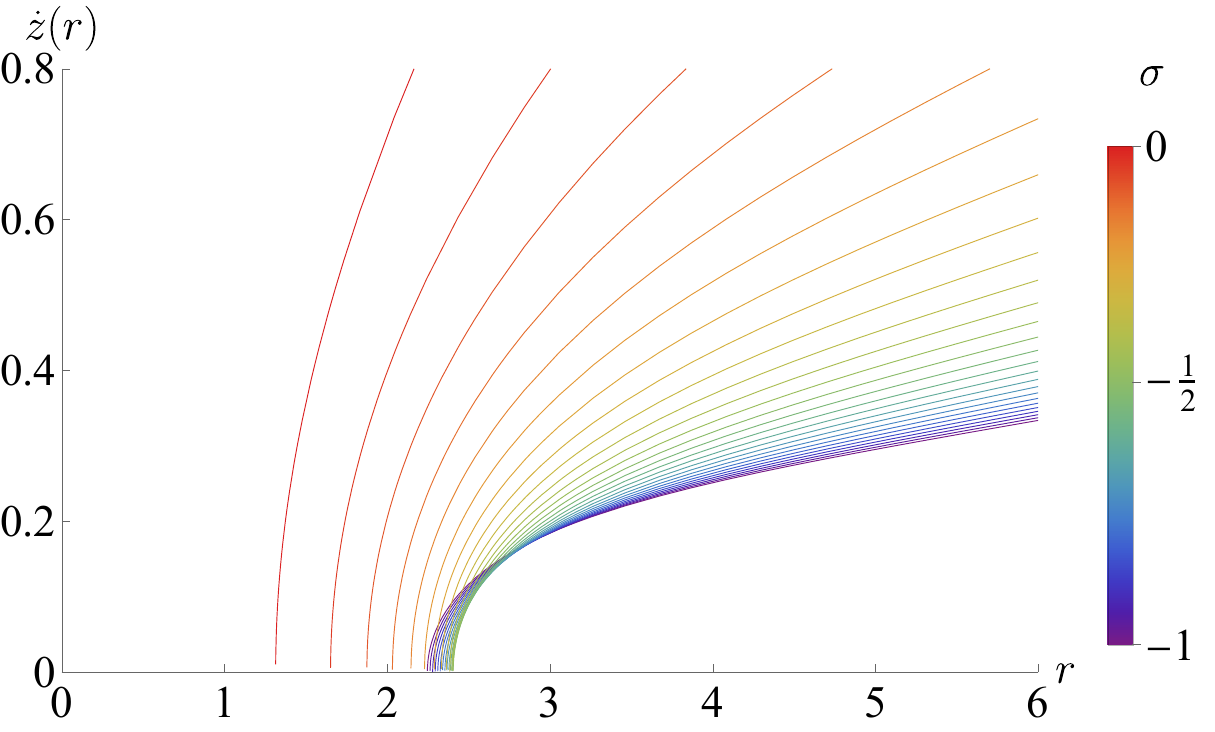}
\caption{$\sigma < 0$}\label{axial velocity plots, Lambda < 0, sigma < 0}
\end{subfigure}%
\begin{subfigure}{.5\textwidth}
\centering
\includegraphics[width=\linewidth]{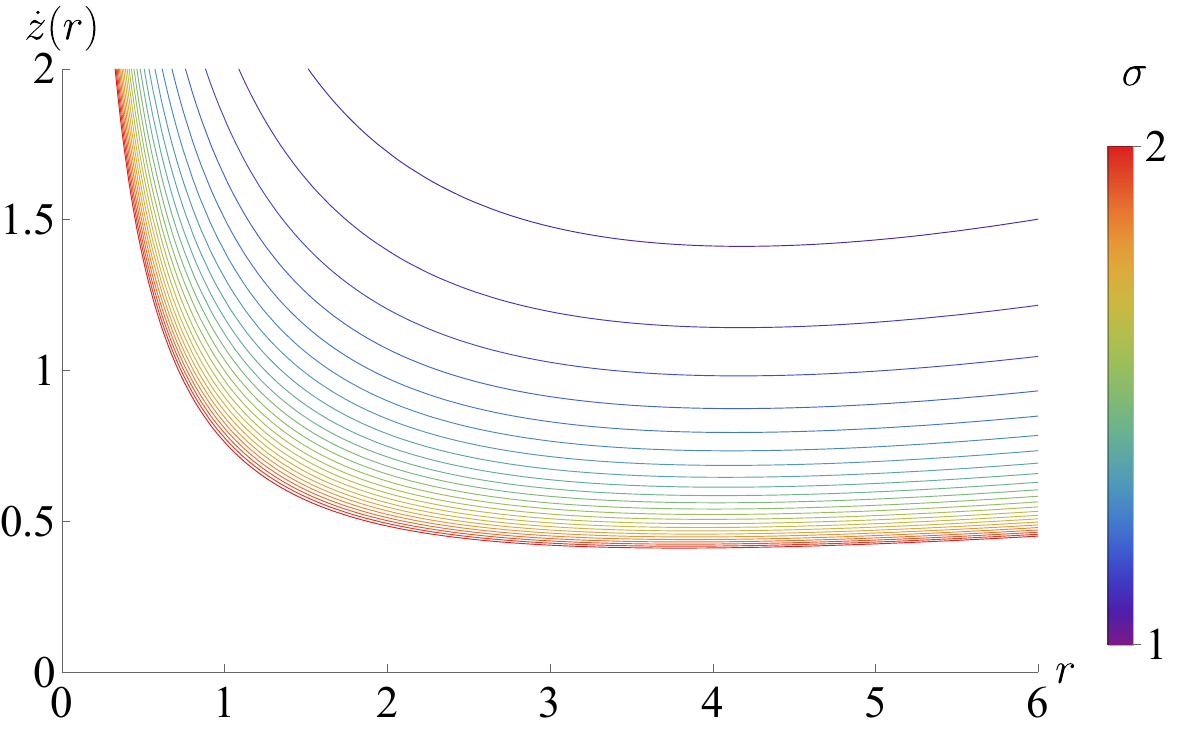}
\caption{$\sigma > 1$}\label{axial velocity plots, Lambda < 0, sigma > 1}
\end{subfigure}
\caption{\label{axial velocity plots, Lambda < 0}The axial velocity for massive particles (\ref{axial geodesic}) with $\Lambda < 0$ (we chose $\Lambda = -0.1$). The curves correspond to various values of $\sigma$ and thus to horizontal cross-sections of \autoref{axial geodesics Lambda < 0}, illustrating the possible values of $\dot{z}(r)$, the corresponding ranges of $r$, and how these depend on $\sigma$. The bottom endpoints of the curves in \autoref{axial velocity plots, Lambda < 0, sigma < 0} correspond to the black curves of static particles in \autoref{axial geodesics Lambda < 0}.}
\end{figure}

Now is the time to ask the question of how it is possible to have these axial paths. Comparing the plots for azimuthal and axial motion, \autoref{angular velocity plots, Lambda > 0} through \autoref{axial velocity plots, Lambda < 0}, we can see they are very similar. As mentioned in \autoref{Proper lengths}, the Einstein equations are local differential equations and thus do not determine the coordinate ranges. Taking further into account the symmetries of the spacetime discussed in \autoref{symmetries}, the idea suggests itself that the axial and azimuthal coordinates were created equal and, in fact, both are of angular nature with a finite range of values and their endpoints identified (there will be an additional conicity parameter for $z$ as well). Once we accept this, we can proceed with the interpretation of the axial motion. If we allow the coordinate $z$ to take values in the range $[0,2\pi)$, implying its angular character and thus introducing a centrifugal force for motion along the $z$-direction, the plots of the axial velocity $\dot{ z}$ can be interpreted in analogy to those of the azimuthal velocity $\dot{\varphi}$.\footnote{We need to introduce a conicity parameter associated with the new angular coordinate in analogy to the parameter $C$ appearing in (\ref{the_metric})---and, likewise, we will set it equal to 1 in the following as well.}

For $\Lambda>0$ and $0<\sigma<1/4$, \autoref{Singularities} and \autoref{proper_lengths} show that the proper circumference $\mathcal{C}_z$ is infinite at $r=0$ and then decreases to zero at $r=\special_radius$ with the centrifugal force pointing toward larger circumferences, that is, from right to left in \autoref{axial velocity plots, Lambda > 0, 0 < sigma < 1/4}, and thus in accordance with our intuition for an attractive singularity at $r=\special_radius$ implied by \autoref{Attractive + repulsive}. This is further corroborated by the fact that the axial circumference $\mathcal{C}_\varphi$ diverges here so that the toroidal surfaces of constant $r$ approach cylindrical surfaces. For $\Lambda>0$ and $\sigma>1$, the proper circumference $\mathcal{C}_z$ vanishes at $r=0$ and then increases to infinity at $r=\special_radius$ with the centrifugal force pointing left to right in \autoref{axial velocity plots, Lambda > 0, sigma > 1}, so that an attractive singularity at $r=0$ implied by \autoref{Attractive + repulsive} is again plausible. The axial circumference $\mathcal{C}_\varphi$ also diverges here so that the toroidal surfaces of constant $r$ approach cylindrical surfaces.

For $\Lambda < 0$ and $\sigma < 0$, the circumference $\mathcal{C}_z$ is an increasing function of $r$ with the centrifugal force pointing from left to right in \autoref{axial velocity plots, Lambda < 0, sigma < 0}. \autoref{Attractive + repulsive} indicates a repulsive singularity at $r=0$ so that the balance is ensured by the cosmological constant at larger radii. This is similar to the situation with azimuthal geodesics and the singularity's ``cylinder of influence'' where it prevents axial geodesics: the cylinder starts with a zero radius for $\sigma = 0$, which then increases to a maximum as $\sigma = -1/2$ and then starts shrinking again as $\sigma \rightarrow - \infty$---see Figure (\ref{Attractive + repulsive}). For $\Lambda < 0$ and $\sigma > 1$, the centrifugal force still points left to right in \autoref{axial velocity plots, Lambda < 0, sigma > 1} and \autoref{Attractive + repulsive} indicates an attractive singularity at $r=0$ providing the balance. Note that a larger $\sigma$ means a weaker attraction here---this is consistent with the fact that $\sigma = 1$ corresponds to a null path.
\subsection{Summary of geodesics}
To make the geodesics easier to grasp, we provide \autoref{the_balance_Lambda_positive} for $\Lambda>0$ and \autoref{the_balance_Lambda_negative} for $\Lambda<0$ explaining where azimuthal and axial geodesics can exist for different values of the parameter $\sigma$---see the table captions for further details.

{\renewcommand{\arraystretch}{1.3} 
\setlength{\tabcolsep}{4pt}
\begin{table}[h]
\begin{centering}
\begin{tabular}{c|cccccc|cccccc|}
\cline{2-13}
 & \multicolumn{6}{c||}{$r=0$} & \multicolumn{6}{c|}{$r=\mathcal{R}$} \\ \hline
\multicolumn{1}{|c||}{\multirow{2}{*}{$\sigma$}} & \multicolumn{2}{c|}{\begin{tabular}[c]{@{}c@{}}Proper\\ lengths\end{tabular}} & \multicolumn{3}{c|}{Force} & \multicolumn{1}{c||}{\multirow{2}{*}{Geod.}} & \multicolumn{1}{c|}{\multirow{2}{*}{Geod.}} & \multicolumn{3}{c|}{Force} & \multicolumn{2}{c|}{\begin{tabular}[c]{@{}c@{}}Proper\\ lengths\end{tabular}} \\ \cline{2-6} \cline{9-13}
\multicolumn{1}{|c||}{} & \multicolumn{1}{c|}{{\color{green}$\mathcal{C}_\varphi$}} & \multicolumn{1}{c|}{{\color{blue}$\mathcal{C}_z$}} & \multicolumn{1}{c|}{{\color{red}$F_s$}} & \multicolumn{1}{c|}{{\color{green}$F_\varphi$}} & \multicolumn{1}{c|}{{\color{blue}$F_z$}} & \multicolumn{1}{c||}{} & \multicolumn{1}{c|}{} & \multicolumn{1}{c|}{{\color{red}$F_s$}} & \multicolumn{1}{c|}{{\color{green}$F_\varphi$}} & \multicolumn{1}{c|}{{\color{blue}$F_z$}} & \multicolumn{1}{c|}{{\color{green}$\mathcal{C}_\varphi$}} & {\color{blue}$\mathcal{C}_z$} \\ \hline\hline
\multicolumn{1}{|c||}{$(1;\infty)$} & \multicolumn{1}{c|}{{\color{green}$\infty$}} & \multicolumn{1}{c|}{{\color{blue}0}} & \multicolumn{1}{c|}{{\color{red}$\leftarrow$}} & \multicolumn{1}{c|}{{\color{green}$\leftarrow$}} & \multicolumn{1}{c|}{{\color{blue}$\rightarrow$}} & \multicolumn{2}{c|}{\diagbox[width=70pt]{\color{blue}axial}{\color{green}azim.}}  & \multicolumn{1}{c|}{{\color{red}$\rightarrow$}} & \multicolumn{1}{c|}{{\color{green}$\leftarrow$}} & \multicolumn{1}{c|}{{\color{blue}$\rightarrow$}} & \multicolumn{1}{c|}{{\color{green}0}} & \multicolumn{1}{c|}{{\color{blue}$\infty$}} \\ \hline
\multicolumn{1}{|c||}{1} & \multicolumn{1}{c|}{{\color{green}$\infty$}} & \multicolumn{1}{c|}{{\color{blue}0}} & \multicolumn{1}{c|}{{\color{red}$\leftarrow$}} & \multicolumn{1}{c|}{{\color{green}$\leftarrow$}} & \multicolumn{1}{c|}{{\color{blue}$\rightarrow$}} & \multicolumn{2}{c|}{\color{blue}axial photons} & \multicolumn{1}{c|}{--} & \multicolumn{1}{c|}{{\color{green}$\leftarrow$}} & \multicolumn{1}{c|}{{\color{blue}$\rightarrow$}} & \multicolumn{1}{c|}{{\color{green}0}} & \multicolumn{1}{c|}{{\color{blue}fin.
}} \\ \hline
\multicolumn{1}{|c||}{$(\frac{1}{2};1)$} & \multicolumn{1}{c|}{{\color{green}$\infty$}} & \multicolumn{1}{c|}{{\color{blue}0}} & \multicolumn{1}{c|}{{\color{red}$\leftarrow$}} & \multicolumn{1}{c|}{{\color{green}$\leftarrow$}} & \multicolumn{1}{c|}{{\color{blue}$\rightarrow$}} & \multicolumn{1}{c||}{--} & \multicolumn{1}{c|}{--} & \multicolumn{1}{c|}{{\color{red}$\leftarrow$}} & \multicolumn{1}{c|}{{\color{green}$\leftarrow$}} & \multicolumn{1}{c|}{{\color{blue}$\leftarrow$}} & \multicolumn{1}{c|}{{\color{green}0}} & \multicolumn{1}{c|}{{\color{blue}0}} \\ \hline
\multicolumn{1}{|c||}{$\frac{1}{2}$} & \multicolumn{1}{c|}{{\color{green}fin.}} & \multicolumn{1}{c|}{{\color{blue}fin.}} & \multicolumn{1}{c|}{{\color{red}$\leftarrow$}} & \multicolumn{1}{c|}{{\color{green}$\leftarrow$}} & \multicolumn{1}{c|}{{\color{blue}$\leftarrow$}} & \multicolumn{1}{c||}{--} & \multicolumn{1}{c|}{--} & \multicolumn{1}{c|}{{\color{red}$\leftarrow$}} & \multicolumn{1}{c|}{{\color{green}$\leftarrow$}} & \multicolumn{1}{c|}{{\color{blue}$\leftarrow$}} & \multicolumn{1}{c|}{{\color{green}0}} & \multicolumn{1}{c|}{{\color{blue}0}} \\ \hline
\multicolumn{1}{|c||}{$(\frac{1}{4};\frac{1}{2})$} & \multicolumn{1}{c|}{{\color{green}0}} & \multicolumn{1}{c|}{{\color{blue}$\infty$}} & \multicolumn{1}{c|}{{\color{red}$\leftarrow$}} & \multicolumn{1}{c|}{{\color{green}$\rightarrow$}} & \multicolumn{1}{c|}{{\color{blue}$\leftarrow$}} & \multicolumn{1}{c||}{--} & \multicolumn{1}{c|}{--}  & \multicolumn{1}{c|}{{\color{red}$\leftarrow$}} & \multicolumn{1}{c|}{{\color{green}$\leftarrow$}} & \multicolumn{1}{c|}{{\color{blue}$\leftarrow$}} & \multicolumn{1}{c|}{{\color{green}0}} & \multicolumn{1}{c|}{{\color{blue}0}} \\ \hline
\multicolumn{1}{|c||}{$\frac{1}{4}$} & \multicolumn{1}{c|}{{\color{green}0}} & \multicolumn{1}{c|}{{\color{blue}$\infty$}} & \multicolumn{1}{c|}{{\color{red}$\leftarrow$}} & \multicolumn{1}{c|}{{\color{green}$\rightarrow$}} & \multicolumn{1}{c|}{{\color{blue}$\leftarrow$}} & \multicolumn{2}{c|}{{\color{green}azim. photons}} & \multicolumn{1}{c|}{--} & \multicolumn{1}{c|}{{\color{green}$\rightarrow$}} & \multicolumn{1}{c|}{{\color{blue}$\leftarrow$}} & \multicolumn{1}{c|}{{\color{green}fin.}} & \multicolumn{1}{c|}{{\color{blue}0}} \\ \hline
\multicolumn{1}{|c||}{$(0;\frac{1}{4})$} & \multicolumn{1}{c|}{{\color{green}0}} & \multicolumn{1}{c|}{{\color{blue}$\infty$}} & \multicolumn{1}{c|}{{\color{red}$\leftarrow$}} & \multicolumn{1}{c|}{{\color{green}$\rightarrow$}} & \multicolumn{1}{c|}{{\color{blue}$\leftarrow$}} & \multicolumn{2}{c|}{\diagbox[dir=SW,width=70pt]{\color{green}azim.}{\color{blue}axial}} & \multicolumn{1}{c|}{{\color{red}$\rightarrow$}} & \multicolumn{1}{c|}{{\color{green}$\rightarrow$}} & \multicolumn{1}{c|}{{\color{blue}$\leftarrow$}} & \multicolumn{1}{c|}{{\color{green}$\infty$}} & \multicolumn{1}{c|}{{\color{blue}0}} \\ \hline
\multicolumn{1}{|c||}{0} & \multicolumn{1}{c|}{{\color{green}0}} & \multicolumn{1}{c|}{{\color{blue}fin.}} & \multicolumn{1}{c|}{--} & \multicolumn{1}{c|}{{\color{green}$\rightarrow$}} & \multicolumn{1}{c|}{{\color{blue}$\rightarrow$}} & \multicolumn{2}{c|}{\color{blue}axial photons} & \multicolumn{1}{c|}{{\color{red}$\rightarrow$}} & \multicolumn{1}{c|}{{\color{green}$\rightarrow$}} & \multicolumn{1}{c|}{{\color{blue}$\leftarrow$}} & \multicolumn{1}{c|}{{\color{green}$\infty$}} & \multicolumn{1}{c|}{{\color{blue}0}} \\ \hline
\multicolumn{1}{|c||}{$(-\frac{1}{2};0)$} & \multicolumn{1}{c|}{{\color{green}0}} & \multicolumn{1}{c|}{{\color{blue}0}} & \multicolumn{1}{c|}{{\color{red}$\rightarrow$}} & \multicolumn{1}{c|}{{\color{green}$\rightarrow$}} & \multicolumn{1}{c|}{{\color{blue}$\rightarrow$}} & \multicolumn{1}{c||}{--} & \multicolumn{1}{c|}{--} & \multicolumn{1}{c|}{{\color{red}$\rightarrow$}} & \multicolumn{1}{c|}{{\color{green}$\rightarrow$}} & \multicolumn{1}{c|}{{\color{blue}$\leftarrow$}} & \multicolumn{1}{c|}{{\color{green}$\infty$}} & \multicolumn{1}{c|}{{\color{blue}0}} \\ \hline
\multicolumn{1}{|c||}{$-\frac{1}{2}$} & \multicolumn{1}{c|}{{\color{green}0}} & \multicolumn{1}{c|}{{\color{blue}0}} & \multicolumn{1}{c|}{{\color{red}$\rightarrow$}} & \multicolumn{1}{c|}{{\color{green}$\rightarrow$}} & \multicolumn{1}{c|}{{\color{blue}$\rightarrow$}} & \multicolumn{1}{c||}{--} & \multicolumn{1}{c|}{--} & \multicolumn{1}{c|}{{\color{red}$\rightarrow$}} & \multicolumn{1}{c|}{{\color{green}$\rightarrow$}} & \multicolumn{1}{c|}{{\color{blue}$\rightarrow$}} & \multicolumn{1}{c|}{{\color{green}fin.}} & \multicolumn{1}{c|}{{\color{blue}fin.}} \\ \hline
\multicolumn{1}{|c||}{$(-\infty;-\frac{1}{2})$} & \multicolumn{1}{c|}{{\color{green}0}} & \multicolumn{1}{c|}{{\color{blue}0}} & \multicolumn{1}{c|}{{\color{red}$\rightarrow$}} & \multicolumn{1}{c|}{{\color{green}$\rightarrow$}} & \multicolumn{1}{c|}{{\color{blue}$\rightarrow$}} & \multicolumn{1}{c||}{--} & \multicolumn{1}{c|}{--} & \multicolumn{1}{c|}{{\color{red}$\rightarrow$}} & \multicolumn{1}{c|}{{\color{green}$\leftarrow$}} & \multicolumn{1}{c|}{{\color{blue}$\rightarrow$}} & \multicolumn{1}{c|}{{\color{green}0}} & \multicolumn{1}{c|}{{\color{blue}$\infty$}} \\ \hline
\end{tabular}
\caption{\label{the_balance_Lambda_positive}Azimuthal and axial geodesics for $\Lambda>0$ as the outcome of the ``forces'' acting on freely falling test particles. The table is divided vertically into two sections with each describing the situation near the singularity at $r=0$ and $r=\mathcal{R}$. The red arrows show which way the nearest singularity pulls the particles. The green and blue arrows point in the direction of the ``centrifugal force'' due to the extrinsic curvature of the azimuthal and axial paths, respectively. It always points toward larger circumferential radii of the corresponding paths, which can be seen in the two columns denoted Proper lengths. A geodesic can only exist where there is balance between the two forces acting upon it: the pull or push of the singularity and the centrifugal force. For instance, the first row informs us that near $r=0$ there can be balance between the leftward pull of the singularity and the rightward pull of the axial centrifugal force---and indeed, axial geodesics exist here as shown in the column headed Geod. Azimuthal geodesics cannot exist here since both forces act in the same direction. There are also some intervals of $\sigma$ for which the balance could, in principle, be established but the pull of the singularity is too strong to overcome at subluminal speeds---these regions are separated from their neighbors by special values of $\sigma$ for which only null geodesics exist. Note that all the information in this table is contained in \autoref{Singularities}, \autoref{Attractive + repulsive}, \autoref{azimuthal geodesics Lambda > 0}, and \autoref{axial  geodesics Lambda > 0}.}
\end{centering}
\end{table}
}
{\renewcommand{\arraystretch}{1.3} 
\setlength{\tabcolsep}{4pt}
\begin{table}[h]
\begin{centering}
\begin{tabular}{c|cccccc|cccccc|}
\cline{2-13}
 & \multicolumn{6}{c||}{$r=0$} & \multicolumn{6}{c|}{$r \rightarrow \infty$} \\ \hline
\multicolumn{1}{|c||}{\multirow{2}{*}{$\sigma$}} & \multicolumn{2}{c|}{\begin{tabular}[c]{@{}c@{}}Proper\\ lengths\end{tabular}} & \multicolumn{3}{c|}{Force} &  \multicolumn{1}{c||}{\multirow{2}{*}{Geod.}} & \multicolumn{1}{c|}{\multirow{2}{*}{Geod.}} & \multicolumn{3}{c|}{Force} & \multicolumn{2}{c|}{\begin{tabular}[c]{@{}c@{}}Proper\\ lengths\end{tabular}} \\ \cline{2-6} \cline{9-13}
\multicolumn{1}{|c||}{} & \multicolumn{1}{c|}{{\color{green}$\mathcal{C}_\varphi$}} & \multicolumn{1}{c|}{{\color{blue}$\mathcal{C}_z$}} & \multicolumn{1}{c|}{{\color{red}$F_s$}} & \multicolumn{1}{c|}{{\color{green}$F_\varphi$}} & \multicolumn{1}{c|}{{\color{blue}$F_z$}} & \multicolumn{1}{c||}{} & \multicolumn{1}{c|}{} & \multicolumn{1}{c|}{{\color{red}$F_{AdS}$}} & \multicolumn{1}{c|}{{\color{green}$F_\varphi$}} & \multicolumn{1}{c|}{{\color{blue}$F_z$}} & \multicolumn{1}{c|}{{\color{green}$\mathcal{C}_\varphi$}} & {\color{blue}$\mathcal{C}_z$} \\ \hline\hline
\multicolumn{1}{|c||}{$(1;\infty)$} & \multicolumn{1}{c|}{{\color{green}$\infty$}} & \multicolumn{1}{c|}{{\color{blue}0}} & \multicolumn{1}{c|}{{\color{red}$\leftarrow$}} & \multicolumn{1}{c|}{{\color{green}$\leftarrow$}} & \multicolumn{1}{c|}{{\color{blue}$\rightarrow$}} & \multicolumn{2}{c|}{\color{blue}axial}  & \multicolumn{5}{c|}{\multirow{7}{*} {\begin{tabular}{@{}c@{}} ${\color{red}F_{AdS}: \leftarrow}$ \\${\color{green}F_\varphi: \rightarrow}$\\${\color{blue}F_z: \rightarrow}$ \\$ {\color{green}\mathcal{C}_\varphi \rightarrow \infty}$ \\${\color{blue}\mathcal{C}_z \rightarrow \infty}$\end{tabular}}} \\ \cline{1-8}
\multicolumn{1}{|c||}{1} & \multicolumn{1}{c|}{{\color{green}$\infty$}} & \multicolumn{1}{c|}{{\color{blue}0}} & \multicolumn{1}{c|}{{\color{red}$\leftarrow$}} & \multicolumn{1}{c|}{{\color{green}$\leftarrow$}} & \multicolumn{1}{c|}{{\color{blue}$\rightarrow$}} & \multicolumn{2}{c|}{\color{blue}axial photons} & \multicolumn{5}{c|}{} \\ \cline{1-8}
\multicolumn{1}{|c||}{$(\frac{1}{4};1)$} & \multicolumn{1}{c|}{{\color{green}0}} & \multicolumn{1}{c|}{{\color{blue}$\infty$}} & \multicolumn{1}{c|}{{\color{red}$\leftarrow$}} & \multicolumn{1}{c|}{{\color{green}$\rightarrow$}} & \multicolumn{1}{c|}{{\color{blue}$\leftarrow$}} & \multicolumn{1}{c||}{--} & \multicolumn{1}{c|}{--} & \multicolumn{5}{c|}{} \\ \cline{1-8}
\multicolumn{1}{|c||}{$\frac{1}{4}$} & \multicolumn{1}{c|}{{\color{green}0}} & \multicolumn{1}{c|}{{\color{blue}$\infty$}} & \multicolumn{1}{c|}{{\color{red}$\leftarrow$}} & \multicolumn{1}{c|}{{\color{green}$\rightarrow$}} & \multicolumn{1}{c|}{{\color{blue}$\leftarrow$}} & \multicolumn{2}{c|}{{\color{green}azim. photons}} & \multicolumn{5}{c|}{} \\ \cline{1-8}
\multicolumn{1}{|c||}{$(0;\frac{1}{4})$} & \multicolumn{1}{c|}{{\color{green}0}} & \multicolumn{1}{c|}{{\color{blue}$\infty$}} & \multicolumn{1}{c|}{{\color{red}$\leftarrow$}} & \multicolumn{1}{c|}{{\color{green}$\rightarrow$}} & \multicolumn{1}{c|}{{\color{blue}$\leftarrow$}} & \multicolumn{2}{c|}{\color{green}azim.} & \multicolumn{5}{c|}{} \\ \cline{1-8}
\multicolumn{1}{|c||}{0} & \multicolumn{1}{c|}{{\color{green}0}} & \multicolumn{1}{c|}{{\color{blue}fin.}} & \multicolumn{1}{c|}{--} & \multicolumn{1}{c|}{{\color{green}$\rightarrow$}} & \multicolumn{1}{c|}{{\color{blue}$\rightarrow$}} & \multicolumn{2}{c|}{\color{blue}axial photons} & \multicolumn{5}{c|}{} \\ \cline{1-8}
\multicolumn{1}{|c||}{$(-\infty;0)$} & \multicolumn{1}{c|}{{\color{green}0}} & \multicolumn{1}{c|}{{\color{blue}0}} & \multicolumn{1}{c|}{{\color{red}$\rightarrow$}} & \multicolumn{1}{c|}{{\color{green}$\rightarrow$}} & \multicolumn{1}{c|}{{\color{blue}$\rightarrow$}} & \multicolumn{2}{c|}{{\color{blue}axial} + {\color{green}azim.}} & \multicolumn{5}{c|}{} \\ \hline
\end{tabular}
\caption{\label{the_balance_Lambda_negative}Azimuthal and axial geodesics for $\Lambda<0$ as the outcome of the ``forces'' acting on freely falling test particles. Since the behavior near $r=0$ is independent of the sign of $\Lambda$, the left part of the table is identical to \autoref{the_balance_Lambda_positive}. The region $r \rightarrow \infty$ is asymptotically AdS with the associated force pointing toward lower $r$'s, and centrifugal forces always pointing toward higher $r$'s for both azimuthal and axial paths. In the range $\sigma \in (1/4;1)$, the singularity is too strong, preventing the existence of geodesics. For further details, please refer to the caption of \autoref{the_balance_Lambda_positive}.}
\end{centering}
\end{table}
}
\section{Symmetries revisited}\label{Symmetries revisited}
Let us now determine the physically relevant interval of $\sigma$. First, recall the symmetries described in \autoref{symmetries} and listed in \autoref{summary_of_symmetries} below.

{\renewcommand{\arraystretch}{2} 
\begin{table}[h]
\begin{centering}
\begin{tabular}{|c|c|c|c|}
\hline
\# & Transformation & $0 \leftrightarrow \special_radius$ & $z \leftrightarrow \varphi$\\ \hline
1 & $\sigma \rightarrow \frac{1-\sigma}{1-4\sigma}$ & + & -- \\ \hline
2 & $\sigma \rightarrow \frac{1-4\sigma}{4(1-\sigma)}$ & + & + \\ \hline
3 & $\sigma \rightarrow \frac{1}{4 \sigma}$ & -- & +\\ \hline
\end{tabular}
\caption{\label{summary_of_symmetries}Transformations preserving the form of the metric (\ref{the_metric}). The column $0 \leftrightarrow \special_radius$ informs us whether the transformation flips the two singularities at $r=0$ and $r = \special_radius$, and the column $z \leftrightarrow \varphi$ tells us whether the transformation switches the axial and azimuthal coordinates. The table applies to $\Lambda>0$ but the last line also works for $\Lambda<0$.}
\end{centering}
\end{table}
}

Assuming $\Lambda>0$, we start from the seed interval $\sigma \in [0;1/4)$ near $r=0$ with azimuthal geodesics but without axial ones (in all these considerations, we refer the reader to \autoref{azimuthal geodesics Lambda > 0} and \autoref{axial geodesics Lambda > 0}). Transformation 1 takes the seed interval to $\sigma \in [1;\infty)$ near $r=\special_radius$ and yields azimuthal geodesics but no axial ones. If we apply transformation 2 instead, we keep $\sigma \in [0;1/4)$ but still move to $r=\special_radius$ and due to the switch $z \leftrightarrow \varphi$, obtain axial geodesics but not azimuthal ones. Applying transformation 3, we end up in the interval $\sigma \in [1,\infty)$ near $r=0$ with axial geodesics but no azimuthal ones. We now argue along the same lines for $\sigma \in (1/4;1/2), \sigma \in [-1/2;0),$ and $\sigma \in (-\infty;-1/2)$ always near $r=0$ with no geodesics possible. Using the three transformations from \autoref{summary_of_symmetries}, we extend these ranges to the entire real axis and to $r=\special_radius$ as well, reproducing the entire geodesic structure of the spacetime as depicted in \autoref{azimuthal_geodesics} and \autoref{axial_geodesics}. This explains why near $r = \special_radius$ there are no azimuthal geodesics for $\sigma \in (-\infty, -1/2)$ and no axial geodesics for $\sigma \in (-1/2;0)$---both are images of the region near $r=0$ and $\sigma \in (1/4;1/2)$ with the attraction of the axis too strong to allow these geodesics. We can apply the same procedure to the case $\Lambda<0$ as well, using the same ranges of $\sigma$ as above but now only the transformation \#3 applies---the asymptotic behavior far from $r=0$ is identical to AdS.

To summarize, the relevant interval of unique $\sigma$ is $(-\infty;1/2]$ as spacetimes with other values of $\sigma$ can be obtained from these via a combination of the transformations in \autoref{summary_of_symmetries}.

\section{Conclusions}\label{conclusions}
In this paper, we have discussed the existence and properties of special geodesics tangent to the spatial coordinate axes in the LC$\Lambda$ spacetime and its symmetries and considered their impact on the interpretation of the spacetime structure and the range of its parameter $\sigma$. Interestingly, it turns out that in addition to the paths that follow $\varphi$, there are also paths that go along $z$. Taking into account the fact that Einstein equations do not distinguish between these two coordinates and do not prescribe their topology, we come to the conclusion that the geodesics are easier to understand if we give the spacetime the toroidal symmetry---that is, if we treat both $\varphi$ and $z$ on the same footing with finite ranges. For $\Lambda>0$ and a general value of $\sigma$, the spacetime features two singularities located at $r=0$ and $r=\special_radius$ with a finite proper distance between them and geodesics navigate the region between them, pulled or pushed in both directions of $r$ depending on their distance from the singularities and the extrinsic curvature of the path, manifesting itself in the form of a centrifugal force. One can imagine the situation as paths between two tire tubes with one embedded within the other so that their axes of rotational symmetry coincide. The tubes are attractive or repulsive depending on the value of $\sigma$. This is only a crude analogy, of course, since we cannot embed the constant-time sections of the spacetime in a 3D Euclidean space, and sometimes the inner tube can be larger than the outer one. At any rate, we can visualize the $\varphi-$ and $z-$geodesics as tracing the great circles on another toroidal surface between the two tubes.

The symmetries of the spacetime enable us to limit the interval of the parameter $\sigma$ appearing in the line element to $(-\infty;1/2]$ since metrics with other values of $\sigma$ are obtained through the symmetry transformations. Additionally, the symmetries further blur the difference between the azimuthal and axial directions.

Both the behavior of geodesics and the symmetries of the spacetime ultimately lead us to the conclusion that the LC$\Lambda$ spacetime makes more sense intuitively if we view it as endowed with toroidal and not cylindrical symmetry.
\section{Acknowledgments}
M.Z. is grateful for the support of grant GACR 22-14791S.
\section*{Bibliography}
\bibliographystyle{iopart-num}
\bibliography{refs}{}

\end{document}